\documentclass[twoside]{article}

\usepackage{psfig}
\usepackage{graphicx}
\usepackage{multicol}
\usepackage{amsfonts}
\usepackage{amssymb}
\usepackage{amsthm}
\usepackage{amsopn}
\usepackage{amsmath}
\usepackage{math}
\usepackage{multirow}
\usepackage{verbatim}
\usepackage{subfigure}
\usepackage{tikz}
\usetikzlibrary{arrows,automata,shapes}

\usepackage{caption}

\newenvironment{Figure}
  {\par\medskip\noindent\minipage{\linewidth}
   }
  {\endminipage\par\medskip}

\newcommand{\captionfonts}{\small}
\makeatletter  
\long\def\@makecaption#1#2{%
 \vskip\abovecaptionskip
 \sbox\@tempboxa{{\captionfonts #1: #2}}%
 \ifdim \wd\@tempboxa >\hsize
   {\captionfonts #1: #2\par}
 \else
   \hbox to\hsize{\hfil\box\@tempboxa\hfil}%
 \fi
 \vskip\belowcaptionskip}
\makeatother   

\newenvironment{Table}
  {\par\medskip\noindent\minipage{\linewidth}
   }
  {\endminipage\par\medskip}

\catcode`\@=11
\long\def\@makefntext#1{
\protect\noindent \hbox to 3.2pt {\hskip-.9pt  
\mbox{\scriptsize$^{{\@thefnmark}}$\hfil}}#1\hfill}	

\def\@makefnmark{\hbox to 0pt{$^{\@thefnmark}$\hss}}	
	
\def\ps@myheadings{%
    \let\@oddfoot\@empty\let\@evenfoot\@empty
    \def\@evenhead{\footnotesize\it\leftmark\hfil}
    \def\@oddhead{\hfil{\footnotesize\it\rightmark}}
    \let\@mkboth\@gobbletwo
    \let\sectionmark\@gobble
    \let\subsectionmark\@gobble
    }

\oddsidemargin=\evensidemargin
\newcounter{sectionc}\newcounter{subsectionc}\newcounter{subsubsectionc}
\renewcommand{\section}[1] {\vspace{14pt}\addtocounter{sectionc}{1}
\setcounter{subsectionc}{0}\setcounter{subsubsectionc}{0}\noindent 
	{\bf\thesectionc. #1}\par\vspace{8pt}}
\renewcommand{\subsection}[1] {\vspace{14pt}\addtocounter{subsectionc}{1}
   \setcounter{subsubsectionc}{0}\noindent 
   {\bf\thesectionc.\thesubsectionc. {\kern1pt \bfit #1}}\par\vspace{8pt}}
\renewcommand{\subsubsection}[1] {\vspace{14pt}
    \addtocounter{subsubsectionc}{1}
	\noindent{\thesectionc.\thesubsectionc.\thesubsubsectionc.
	{\kern1pt \it #1}}\par\vspace{8pt}}
\newcommand{\nonumsection}[1] {\vspace{14pt}\noindent{\bf #1}
	\par\vspace{8pt}}

\newcounter{appendixc}
\newcounter{subappendixc}[appendixc]
\newcounter{subsubappendixc}[subappendixc]
\renewcommand{\thesubappendixc}{\Alph{appendixc}.\arabic{subappendixc}}
\renewcommand{\thesubsubappendixc}
	{\Alph{appendixc}.\arabic{subappendixc}.\arabic{subsubappendixc}}

\renewcommand{\appendix}[1] {\vspace{14pt}
        \refstepcounter{appendixc}
        \setcounter{figure}{0}
        \setcounter{table}{0}
        \setcounter{lemma}{0}
        \setcounter{theorem}{0}
        \setcounter{corollary}{0}
        \setcounter{definition}{0}
        \setcounter{equation}{0}
        \renewcommand{\thefigure}{\Alph{appendixc}.\arabic{figure}}
        \renewcommand{\thetable}{\Alph{appendixc}.\arabic{table}}
        \renewcommand{\theappendixc}{\Alph{appendixc}}
        \renewcommand{\thelemma}{\Alph{appendixc}.\arabic{lemma}}
        \renewcommand{\thetheorem}{\Alph{appendixc}.\arabic{theorem}}
        \renewcommand{\thedefinition}{\Alph{appendixc}.\arabic{definition}}
        \renewcommand{\thecorollary}{\Alph{appendixc}.\arabic{corollary}}
        \renewcommand{\theequation}{\Alph{appendixc}.\arabic{equation}}
        \noindent{\bf Appendix \theappendixc #1}\par\vspace{5pt}}
\newcommand{\subappendix}[1] {\vspace{14pt}
        \refstepcounter{subappendixc}
        \noindent{\bf Appendix \thesubappendixc. {\kern1pt \bfit #1}}
	\par\vspace{8pt}}
\newcommand{\subsubappendix}[1] {\vspace{14pt}
        \refstepcounter{subsubappendixc}
        \noindent{\rm Appendix \thesubsubappendixc. {\kern1pt \it #1}}
	\par\vspace{8pt}}

\newcommand{\textlineskip}{\baselineskip=13pt}
\newcommand{\smalllineskip}{\baselineskip=10pt}

\newcommand{\copyrightheading}[1]
	{\vspace*{-2.5cm}\smalllineskip{\flushleft
	{\footnotesize arXiv preprint, #1}
	 }}

\def\abstracts#1#2{{
	\centering{\begin{minipage}{5.8in}\small\baselineskip=11pt
	\parindent=0pc #1\par 
	\parindent=2pc #2
	\end{minipage}}\par}}

\renewenvironment{thebibliography}[1]		
	{\small\baselineskip=11pt
	 \frenchspacing
         \begin{list}{\arabic{enumi}.}
         {\usecounter{enumi}
         \setlength{\parsep}{0pt}    
         \setlength{\leftmargin}{12.7pt}
         \setlength{\rightmargin}{0pt}
         \setlength{\itemsep}{0pt} \settowidth
         {\labelwidth}{#1.}\sloppy}}
         {\end{list}}

\newcounter{itemlistc}
\newcounter{romanlistc}
\newcounter{alphlistc}
\newcounter{arabiclistc}

\newcommand{\fcaption}[1]{
        \refstepcounter{figure}
        \setbox\@tempboxa = \hbox{\small Fig.~\thefigure. #1}
        \ifdim \wd\@tempboxa > 5in
           {\begin{center}
        \parbox{5in}{\small\baselineskip=11pt Fig.~\thefigure. #1}
            \end{center}}
        \else
             {\begin{center}
             {\small Fig.~\thefigure. #1}
              \end{center}}
        \fi}

\newcommand{\tcap}[1]{
        \refstepcounter{table}
        \setbox\@tempboxa = \hbox{\small Table~\thetable. #1}
        \ifdim \wd\@tempboxa > 3.15in
           {\begin{center}
        \parbox{3.15in}{\small\smalllineskip 
	    Table~\thetable. #1}
            \end{center}}
        \else
             {\begin{center}
             {\eightpoint Table~\thetable. #1}
              \end{center}}
        \fi}

\def\@citex[#1]#2{\if@filesw\immediate\write\@auxout
	{\string\citation{#2}}\fi
\def\@citea{}\@cite{\@for\@citeb:=#2\do
	{\@citea\def\@citea{,}\@ifundefined
	{b@\@citeb}{{\bf ?}\@warning
	{Citation `\@citeb' on page \thepage \space undefined}}
	{\csname b@\@citeb\endcsname}}}{#1}}

\newif\if@cghi
\def\cite{\@cghitrue\@ifnextchar [{\@tempswatrue
	\@citex}{\@tempswafalse\@citex[]}}
\def\citelow{\@cghifalse\@ifnextchar [{\@tempswatrue
	\@citex}{\@tempswafalse\@citex[]}}
\def\@cite#1#2{{$\null^{#1}$\if@tempswa\typeout
	{IJCGA warning: optional citation argument 
	ignored: `#2'} \fi}}

\def\pmb#1{\setbox0=\hbox{#1}
	\kern-.025em\copy0\kern-\wd0
	\kern.05em\copy0\kern-\wd0
	\kern-.025em\raise.0433em\box0}

\def\fnm#1{$^{\mbox{\scriptsize #1}}$}		
\def\fnt#1#2{\footnotetext{\kern-.3em
	{$^{\mbox{\scriptsize #1}}$}{#2}}}

\font\tenit=cmti10 

\font\bfit=cmbxti10 at 10pt

\font\eightit=cmti8

\def\itlatex{\tenit L\kern-.30em\raise.4ex\hbox{\eightit A}\kern-.14em 
T\kern-.1667em\lower.7ex\hbox{E}\kern-.125em X} 

\def\bsc{{\sc a\kern-7pt\sc a}}
\def\bflatex{\bf L\kern-.30em\raise.3ex\hbox{\bsc}\kern-.18em
T\kern-.1667em\lower.7ex\hbox{E}\kern-.125em X} 

\def\qed{\hbox{${\vcenter{\vbox{			
   \hrule height 0.4pt\hbox{\vrule width 0.4pt height 6pt
   \kern5pt\vrule width 0.4pt}\hrule height 0.4pt}}}$}}

\newcommand{\black}[1]{\textcolor{black}{#1}}

\begin{document}
\setlength{\textheight}{8.78truein}     

\thispagestyle{empty}

\markboth{ACOUSTIC SPACE LEARNING FOR SOUND SOURCE SEPARATION AND LOCALIZATION}
{ACOUSTIC SPACE LEARNING FOR SOUND SOURCE SEPARATION AND LOCALIZATION}

\textlineskip
\setcounter{page}{1}

\copyrightheading{February 2014} 

\vspace*{1.05truein}


\centerline{\large\bf \black{ACOUSTIC SPACE LEARNING FOR SOUND-SOURCE SEPARATION}}
\vspace*{0.08truein}
\centerline{\large\bf \black{AND LOCALIZATION ON BINAURAL MANIFOLDS}}

\vspace*{0.45truein}
\centerline{ANTOINE DELEFORGE}
\vspace*{0.0215truein}
\centerline{\it INRIA Grenoble Rh\^{o}ne-Alpes, 655 Avenue de l'Europe}
\baselineskip=11pt
\centerline{\it Saint-Ismier, 38334, France}
\centerline{\it E-mail: antoine.deleforge@inria.fr}
\vspace*{14pt}
\centerline{FLORENCE FORBES}
\vspace*{0.0215truein}
\centerline{\it INRIA Grenoble Rh\^{o}ne-Alpes, 655 Avenue de l'Europe}
\centerline{\it Saint-Ismier, 38334, France}
\centerline{\it E-mail: florence.forbes@inria.fr}
\vspace*{14pt}
\centerline{RADU HORAUD}
\vspace*{0.0215truein}
\centerline{\it INRIA Grenoble Rh\^{o}ne-Alpes, 655 Avenue de l'Europe}
\centerline{\it Saint-Ismier, 38334, France}
\centerline{\it E-mail: radu.horaud@inria.fr}



\vspace*{0.29truein}
\abstracts{In this paper we address the problems of modeling the acoustic space generated by a full-spectrum sound source and of using the learned model for the localization and separation of multiple sources that simultaneously emit sparse-spectrum sounds. We lay theoretical and methodological grounds \black{in order} to introduce the \textit{binaural manifold} paradigm. We perform an in-depth study of the latent low-dimensional structure of the high-dimensional interaural spectral data, based on a \black{corpus} recorded with a human-like audiomotor robot head. A non-linear dimensionality reduction technique is used to show that these data lie on a two-dimensional (2D) smooth manifold parameterized by the motor states of the listener, or equivalently, the sound source directions. We propose a \textit{probabilistic piecewise affine mapping} model (PPAM) specifically designed to deal with high-dimensional data  exhibiting an intrinsic piecewise linear structure. We derive a closed-form expectation-maximization (EM) procedure for estimating the model parameters, followed by Bayes inversion for obtaining the full posterior density function of a sound source direction. We extend this solution to deal with missing data and redundancy in real world spectrograms, and hence for 2D localization of natural sound sources such as speech. We further generalize the model to the challenging case of multiple sound sources and we propose a variational EM framework. The associated algorithm, referred to as \textit{variational EM for source separation and localization} (VESSL) yields a Bayesian estimation of the 2D locations and time-frequency masks of all the sources. Comparisons of the proposed approach with several existing methods reveal that the combination of acoustic-space learning with Bayesian inference \black{enables} our method to outperform state-of-the-art methods.
%
}{}
\vspace*{10pt}\textlineskip

\begin{multicols}{2}

\section{Introduction}
The \black{remarkable abilities of humans} to localize one or several sound sources and to identify their content
from the perceived acoustic signals have been intensively studied 
in psychophysics\cite{Blauert97}, computational auditory analysis\cite{WangBrown2006},
and more recently in the emerging field of \textit{robot hearing}\cite{DeleforgeHoraud12a}.
A classical example \black{which} illustrates the difficulty of understanding these human skills, is the well known \textit{cocktail party effect} 
introduced by Cherry\cite{Cherry53} that still challenges today's methods\cite{HaykinChen2005}: How \black{are} listeners able to decipher speech in the presence
of other sound sources, including competing talkers? While human listeners solve this problem routinely and effortlessly, this is still a challenge in computational audition.

There is behavioral and physiological evidence that human listeners
use \textit{binaural cues} in order to estimate the direction of a
sound source\cite{Rayleigh1907} and that sound localization plays an important role for solving the cocktail party problem\cite{Middlebrooks91,WangBrown2006}. Two binaural cues seem to play an essential role, namely the
interaural level difference (ILD) and the interaural time difference
(ITD), or its spectral equivalent the interaural phase difference (IPD).
Both ILD and IPD are known to be
subject-dependent and frequency-dependent cues, due to the so-called \textit{head related transfer function} (HRTF) generated by the shape of the head, pinna and torso,
which filters signals arriving at each eardrum. \black{These complex shapes induce a non-linear dependency of the HRTFs on the sound source direction\cite{Blauert97,WangBrown2006}}. It is known that the spatial
information provided by interaural-difference cues within a
restricted band of frequency is spatially ambiguous, particularly
along a roughly vertical and front/back dimension\cite{Middlebrooks91}. This suggests that humans and mammals make use of full spectrum information for 2D sound source localization\cite{WoodworthSchlosberg65,HofmanVanOpstal98}.
This is confirmed by biological models of the auditory system hypothesizing the existence of neurons dedicated to the computation of interaural cues in specific frequency bands\cite{WangBrown2006}.

A lot of computational techniques exist to extract ITD, ILD and IPD from binaural recordings, either in the time domain using cross-correlation\cite{LiuWang10,Alameda-EUSIPCO2012}, or in the time-frequency domain using Fourier analysis\cite{MandelEllisJebara07,DeleforgeHoraud12b} or gammatone filters\cite{WoodruffWang12}.
However, the problem of localizing and/or separating several sound sources remains a challenge in computational auditory scene analysis, for
several reasons. Firstly, the mapping from interaural cues to sound-source positions
is usually unknown, complex and non-linear due to the
transfer function of microphones which cannot be easily
modeled. Secondly, auditory data are corrupted by noise and
reverberations. Thirdly, an interaural value at a given frequency is relevant only
if the source is actually emitting at that frequency: Natural
sounds such as speech are known to be extremely sparse, with often
80\% of the frequencies actually missing at a given time. Finally,
when several sources emit simultaneously, the assignment of a
time-frequency point to one of the sources
is \black{hard to estimate}. The first problem, \textit{i.e.}, mapping audio cues to source positions, is central: Yet, it has received little attention in computational audition. Most existing approaches approximate this
mapping based on simplifying assumptions, such as direct-path
source-to-microphone propagation\cite{YilmazRickard04},
 a sine interpolation of ILD data from a human HRTF
dataset\cite{VisteEvangelista03}, or a spiral ear model\cite{KullaibAlmuallaVernon09}. These simplifying assumptions are often not valid in real world
conditions. Following this view, accurately modeling a real world binaural system would require a prohibitively high number of parameters including \black{the exact shape and acoustic properties of the recording device and of the room}, which are unaccessible in practice. Due to this difficulty, the vast majority of current binaural sound localization
approaches mainly focus on a rough estimation of a frontal azimuth angle, or
\textit{one-dimensional} (1D) localization\cite{LiuWang10,MandelEllisJebara07,WoodruffWang12,VisteEvangelista03},
and very few perform 2D localization\cite{KullaibAlmuallaVernon09}. Alternatively, some approaches\cite{KeyrouzMaierDiepold07,HornsteinLopesSantosVictorLacerda2006,DeleforgeHoraud12b}
bypass the explicit mapping model and perform 2D localization
by exhaustive search in a large HRTF look-up table associating source directions to interaural spectral cues. However, this
process is unstable and hardly scalable in practice as the number
of required associations yields prohibitive memory and
computational costs.

On the other hand, a number of psychophysical studies have suggested that the ability of localizing sounds is learned at early stages of development in humans and mammals\cite{Hofman98,WrightZhang06,Aytekin08}. That is, the link between auditory features and source locations might not be hardcoded in the brain but rather learned from experience. The sensorimotor theory, originally laid by Poincar\'{e}\cite{Poincare29} and more recently investigated by O'Regan\cite{oregan_noe}, suggests that experiencing the sensory consequences of voluntary motor actions is necessary for an organism to learn the perception of space. For example, Held and Hein\cite{held_hein} showed that neo-natal kittens deprived \black{of} the ability of moving while seeing could not develop vision properly. Most notably, Aytekin et al.\cite{Aytekin08} proposed a sensorimotor model of sound source localization using HRTF datasets of bats and humans. In particular, they argue that biological mechanisms \black{could learn} sound localization based solely on acoustic inputs and their relation to motor states.

In this article, we get inspiration from these psychological studies to propose a \textit{supervised learning} computational paradigm for multiple sound source separation and localization. In other words, the acoustic properties of a system are first learned during a \textit{training phase}. A key element of this study is the existence of \textit{binaural manifolds}, which are defined and detailed in this article. Their existence is asserted through an in-depth study of the intrinsic structure of high-dimensional interaural spectral cues, based on real world recordings with a human-like auditory robot head. A non-linear dimensionality reduction technique is used to show that these cues lie on a two-dimensional (2D) smooth manifold parameterized by the motor states of the system, or equivalently, the sound source directions. With this key observation in mind, we propose a \textit{probabilistic piecewise affine mapping} model (PPAM) specifically designed to deal with high-dimensional data presenting such a locally linear structure. We explain how the model parameters can be learned using a closed-form expectation-maximization (EM) procedure, and how Bayes inversion can be used to obtain the full posterior density function of a sound source direction given a new auditory observation. We also show how to extend this inversion to deal with missing data and redundancy in real world spectrograms, and hence for two-dimensional localization of natural sound sources such as speech. We further extend this inversion to the challenging case of multiple sound sources emitting at the same time, by proposing a variational EM (VEM) framework. The associated algorithm, referred to as \textit{variational EM for source separation and localization} (VESSL) yields a Bayesian estimation of the 2D locations and time-frequency masks of all the sources. We introduced some parts of this paradigm in previous work\cite{DeleforgeHoraud12c,DeleforgeHoraud13}, and we propose in this article a unified, more detailed and in-depth presentation of the global methodology.

The remainder of this article is organized as follows.
Section~1.1~
provides a brief literature overview of sound source separation and space mapping.
Section~2~
describes how interaural cues are computed in our framework.
Section~3~
presents the audio-motor dataset specifically recorded for this study.
Section~4~
describes the manifold learning method used to prove the existence of smooth binaural manifolds parameterized by source directions.
Section~5~
describes the PPAM model, its inference using EM, as well as the associated inverse mapping formula extended to missing and redundant observations in spectrograms.
Section~6~
further \black{extends} the inversion to the case of multiple sound sources with the \black{VESSL algorithm}.
Section~7~
shows detailed performance of PPAM and VESSL in terms of mapping, sound separation and sound localization under various conditions. Results are compared to state-of-the-art methods.
Finally, Section~8~
concludes and provides directions for future work.

\subsection{Related work}
\label{sec:related}
\noindent
\textbf{Sound Source Separation.}
The problem of sound source separation has been thoroughly studied in the last decades and several interesting
approaches have been proposed. Some methods\cite{Roweis00,BensaidAntony10}
achieve separation with a single microphone, based on known
acoustic properties of speech signals, and are therefore limited to a specific type of input.
Other techniques such as independent component analysis (ICA)\cite{ComonJutten10} require as many microphones as the number of sources.
A recently explored approach is to use Gaussian complex models for spectrograms and estimate source parameters using an expectation-maximization algorithm\cite{Duong10}.
However, these methods are known to be very sensitive to initialization due to the large dimensionality of the parameter space, and often rely on external
knowledge or other sound source separation algorithms for initialization.

Another category of methods \black{uses} binaural localization cues combined with \textit{time-frequency masking} for
source separation\cite{VisteEvangelista03,KeyrouzMaierDiepold07,MandelWeissEllis10}.
Time-frequency masking, also called binary masking,
allows the separation of \black{more sources than microphones} from a mixed signal, with the assumption that a single
source is active at every time-frequency point.
This is referred to as the \textit{W-disjoint orthogonality} assumption\cite{YilmazRickard04} and it
has been shown to hold, in general, for simultaneous speech signals.
Recently, Mandel et al.\cite{MandelWeissEllis10} proposed a probabilistic model
for multiple sound source localization and separation based on interaural spatial cues and binary masking. For each
sound source, a binary mask and a discrete distribution over interaural time delays is provided.
This can be used to approximate the frontal azimuth angle of the sound source using a direct-path sound propagation model,
if the distance between the microphones is known.

\noindent
\textbf{Space Mapping.}
The task of learning a mapping between two spaces can be summarized as follows: if we are given a set of training couples
$\{(\xvect_n,\yvect_n)\}_{n=1}^N\subset \mathbb{R}^L\times \mathbb{R}^D$,
how can we obtain a relationship between $\mathbb{R}^L$ and $\mathbb{R}^D$ such that given a new observation in one space,
its associated point in the other space is deduced?
This problem has been extensively studied in machine learning, and offers a broad range of applications.
In this study, we are interested in mapping a high-dimensional space to a low dimensional space, where the relationship
between the two spaces is approximately locally linear. Indeed, it is showed in Section 4
that high dimensional spectral interaural
data lie on a smooth Riemanian manifold, which by definition is locally homeomorphic to an Euclidean space.
High-to-low-dimensional mapping
problems are often solved in two steps, \textit{i.e.}, dimensionality
reduction followed by regression. Methods falling into this category are partial
least-squares\cite{rosipal2006overview}, sliced inverse
regression (SIR)\cite{Li91} and kernel SIR\cite{wu2008kernel}, but they are not designed to model local linearity.
An attractive probabilistic approach for local linearity
is to use a mixture of locally linear sub-models. A number of methods have explored this
paradigm in the Gaussian case, such as mixture of linear regressors\cite{deVeaux89},
mixture of local experts\cite{XuJordanHinton95} and methods based on joint Gaussian mixture models\cite{KainMacon98,StylianouMoulines98,TodaBlackTokuda08,QiaoMinematsu09} (GMM), \black{that were} recently unified
in a common supervised and unsupervised mapping framework\cite{DeleforgeHoraud13b}.
Joint GMM has particularly been used in audio applications such as text-to-speech synthesis, voice conversion or articulatory-acoustic mapping systems. In Section 5,
we propose a variant of the mixture of local experts model\cite{XuJordanHinton95} specifically adapted to locally linear and high-to-low dimensional mapping with an \black{attractive} geometrical interpretation. The concept of \textit{acoustic space learning} introduced in this article, \textit{i.e.}, the idea of learning a mapping from interaural cues to 2D source positions and exploiting this mapping for sound source separation and localization does not seem to have been explored in the past.

\section{Extracting Spatial Cues from Sounds}
\label{sec:cues}
\black{To localize} sound sources, one \black{needs} to find a representation of sounds that 1) \black{is independent} of the sound source \textit{content}, \textit{i.e.}, the emitted signal and 2) contains discriminative spatial information. These features can be computed in the time domain, but contain richer information when computed for different frequency channels, \textit{i.e.}, in the time-frequency domain. This section presents the time-frequency representation and spatial features used in this study.

\subsection{Time-Frequency Representation}
\label{subsec:spectrogram}
A time-frequency representation can be obtained either \black{using Gammatone filter banks\cite{WoodruffWang12}, which are mimicking human auditory representation, or short-term Fourier transform (STFT) analysis\cite{MandelEllisJebara07,DeleforgeHoraud12b}}. The present study uses STFT as it is more directly applicable to source separation through binary-masking, as addressed in
Section~6.~
First, the complex-valued spectrograms associated with the two microphones are computed with a 64 ms time-window and 8 ms \black{hop}, yielding $T=126$ frames for a 1 s signal. \black{These values proved to be a good compromise between computational time and time-frequency resolution}. Since sounds are recorded at 16,000 Hz, each time window contains 1,024 samples which are transformed into $F=512$ complex Fourier coefficients associated to positive frequency channels between 0 and 8,000 Hz. For a binaural recording made in the presence of a sound source located at $\xvect$ in a listener-centered coordinate frame, we denote with $\{s^{(\textrm{S})}_{ft}\}_{f,t=1}^{F,T}$ the complex-valued spectrogram emitted by the source, and with $\{s^{(\textrm{L})}_{ft}\}_{f,t=1}^{F,T}$ and $\{s^{(\textrm{R})}_{ft}\}_{f,t=1}^{F,T}$ the left and right perceived spectrograms.

\subsection{Interaural Spectral Cues}
\label{subsec:int_cues}
The HRTF model provides a relationships between the spectrogram emitted from source position $\xvect$ and the perceived spectrograms:
\begin{equation}
\label{eq:hrtf_kodel}
\left\{
\begin{array}{l}
s^{(\textrm{L})}_{ft} = h_f^{(\textrm{L})}(\xvect) \; s^{(\textrm{S})}_{ft} \\
s^{(\textrm{R})}_{ft} = h_f^{(\textrm{R})}(\xvect) \; s^{(\textrm{S})}_{ft}.
\end{array}
\right.
\end{equation}
where $\hvect^{(\textrm{L})}$ and $\hvect^{(\textrm{R})}$ denote the left and right non-linear HRTFs.
The \textit{interaural transfer function} (ITF) is defined by the ratio between the two HRTFs, \textit{i.e.}, 
$I_f(\xvect)=h_f^{(\textrm{R})}(\xvect)/h_f^{(\textrm{L})}(\xvect)\in\mathbb{C}$. The interaural spectrogram is defined by 
$\hat{I}_{ft} = s^{(\textrm{R})}_{ft} / s^{(\textrm{L})}_{ft}$, so that $\hat{I}_{ft} = I_f(\xvect)$. This way, $\hat{I}_{ft}$ does not depend on the emitted spectrogram value $s^{(S)}_{ft}$ but only on the emitting source position $\xvect$. However, this equality holds only if the source is emitting at $(f,t)$ (\textit{i.e.}, $s^{(\textrm{S})}_{ft}\ne0$). Since natural sounds have a null acoustic level at most time-frequency bins, associated interaural spectrograms will be very sparse, \textit{i.e.}, they will have \black{many} \textit{missing values}. To characterize missing interaural spectrogram values, we introduce the binary variables $\chi_{ft}$ so that $\chi_{ft}=0$ if the value at $(f,t)$ is missing and $\chi_{ft}=1$ otherwise. They can be determined using a threshold on left and right spectral powers $|s^{(\textrm{L})}_{ft}|^2$ and $|s^{(\textrm{R})}_{ft}|^2$.
We define the \textit{ILD spectrogram} $\alphavect$ and the \textit{IPD spectrogram} $\phivect$ as the log-amplitude and phase of the interaural spectrogram $\hat{I}_{f,t}$:
\begin{equation}
\left\{
\begin{array}{l}
\alpha_{ft}=20\log|\hat{I}_{ft}| \in \mathbb{R}, \\
\phi_{ft}=\exp (j \arg(\hat{I}_{ft})) \in \mathbb{C}.
\end{array}
\right.
\end{equation}
The phase difference is expressed in the complex domain, or equivalently $\mathbb{R}^2$, to avoid problems due to phase circularity. This way, two nearby phase values will be nearby in terms of Euclidean distance. In the particular case of a sound source emitting white noise from $\xvect$, we have \black{$\chi_{ft}=1$} for all $(f,t)$, \textit{i.e.}, the sound source is emitting at all $(f,t)$ points. One can thus compute the temporal means $\bar{\alphavect}\in\mathbb{R}^F$ and $\bar{\phivect}\in\mathbb{R}^{2F}$ of ILD and IPD spectrograms. These vectors will be referred to as the \textit{mean interaural vectors} associated \black{with} $\xvect$.
The well established duplex theory\cite{Middlebrooks91} suggests that ILD cues are mostly used at high frequencies (above 2  kHz) while ITD (or IPD) cues are mostly used at low frequencies (below 2 kHz) in humans. Indeed, ILD values are similar at low frequencies because the HRTF can be neglected, and the phase difference becomes very unstable with respect to the source position at high frequencies. To account for these phenomena, the initial binaural cues are split into two distinct vectors, namely 
the mean \textit{low}-ILD and \textit{high}-ILD and the mean \textit{low}-IPD and \textit{high}-IPD vectors, where \textit{low} corresponds to frequency channels between 0 and 2 kHz and \textit{high} corresponds to frequency channels between 2 kHz and 8 kHz. Frequency channels below 300 Hz were also removed in \textit{low}-IPD vectors as they did not \black{prove} to contain significant information, due to a high amount of noise in this frequency range.

\section{The CAMIL Dataset: Audiomotor Sampling of the Acoustic Space}
\label{sec:dataset}
\begin{figure*}[t!]
    \centering
    \includegraphics[width = 0.9\linewidth,clip=,keepaspectratio]{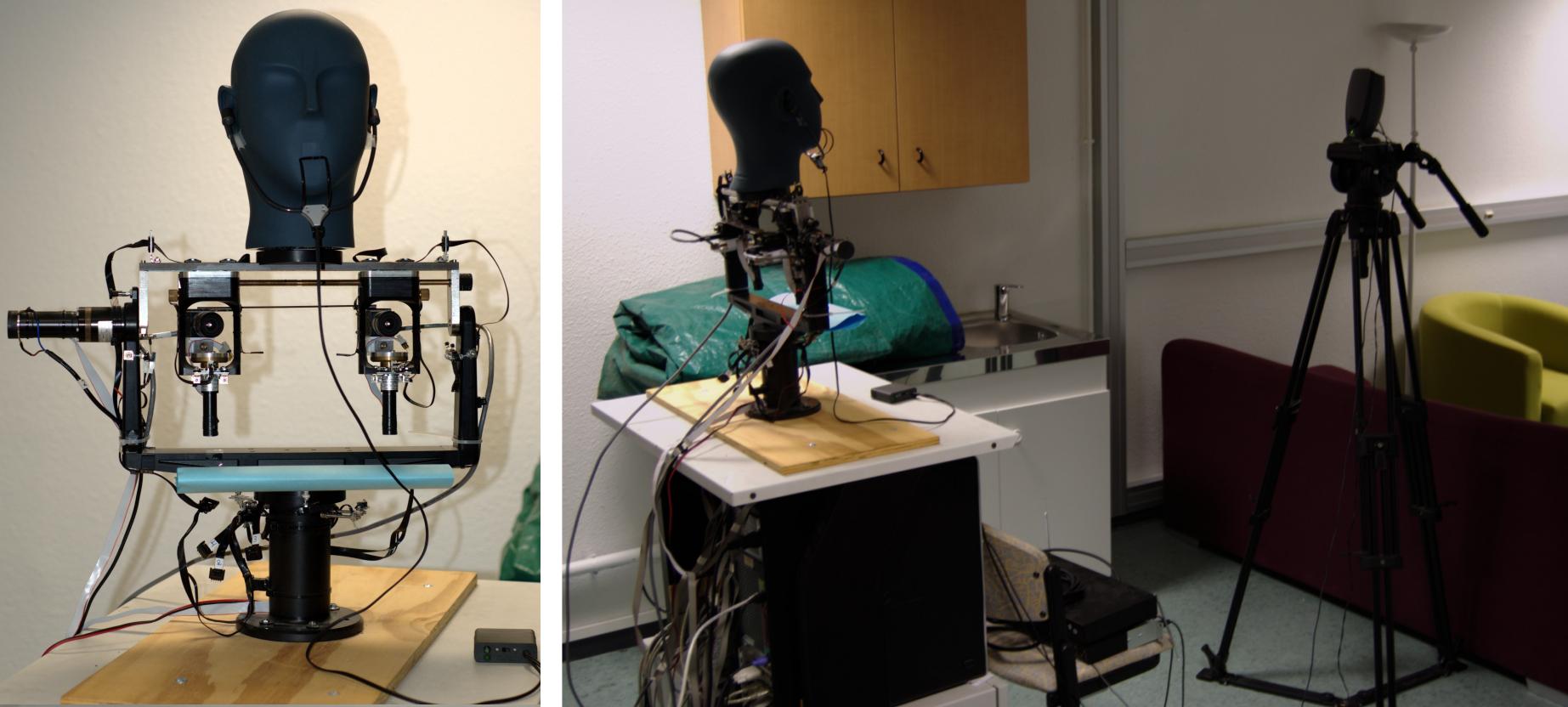}
    \caption{\label{fig:exp_photos}\small{A binaural dummy head is placed
      onto a motor system which can perform precise and reproducible
      pan and tilt motions (left). The emitter (a loud-speaker) is placed in
      front of the robot (right).}}
\end{figure*}

We developed a technique to gather a large number of interaural vectors associated \black{with} source positions, 
in an entirely unsupervised and automated way.
Sound acquisition is performed with a Sennheiser MKE 2002 acoustic dummy-head
linked to a computer via a Behringer ADA8000 Ultragain
Pro-8 digital external sound card. The head is mounted onto a
robotic system with two rotational degrees of freedom: a pan motion and a tilt motion (see Fig.~\ref{fig:exp_photos}).
This device is specifically designed to achieve precise and reproducible
movements\fnm{a}\fnt{a}{Full details on the setup at http://team.inria.fr/perception/popeye/}. The emitter -- a
loud-speaker -- is placed at approximately 2.7 meters ahead of the
robot, as shown on Fig.~\ref{fig:exp_photos}. The loud-speaker's input and the microphones' outputs are
handled by two synchronized sound cards in order to simultaneously
record and play. All the experiments are carried out in real-world conditions,
\textit{i.e.}, in a room with natural reverberations and background noise due to computer fans.

Rather than placing the emitter at known 3D locations around the robot, it is
kept in a fixed reference position while the robot records emitted sounds
from different motor states. Consequently, a sound source direction is
directly associated \black{with} a pan-tilt motor state $(\psi,\theta)$.
\black{A great advantage of the audio-motor method is that it allows to obtain a very dense sampling of almost the entire acoustic space. The overall training is fully-automatic and takes around 15 hours, which could not be done manually. However, it also presents a limitation: A recording made at a given motor-state only approximates what would be perceived if the source was actually placed at the corresponding relative position in the room. This approximation holds only if the room has relatively few asymmetries and reverberations, which might not be the case in general. Note that when this is the case, a sound localization system trained with this dataset could be used to directly calculate the head movement pointing toward an emitting sound source. This could be done without needing inverse kinematics, distance between microphones or any other parameters. Alternatively, one could learn the room acoustics together with the system acoustics using, \textit{e.g.}, visual markers related to an emitter successively placed in different positions around the static head.} Interestingly, the remainder of this study does not depend on the specific way source-position-to-interaural-cues associations are gathered.

\black{Recordings are made from $10,800$ uniformly spread motor states covering the entire reachable motor-state space of the robot:}
$180$ pan rotations $\psi$ in the range $[-180^{\circ},180^{\circ}]$ (left-right) and
$60$ tilt rotations $\theta$ in the range
$[-60^{\circ},60^{\circ}]$ (top-down). Hence, the source location spans a $360^{\circ}$ azimuth
range and a $120^{\circ}$ elevation range in the robot's frame, with $2^\circ$ between each source direction. There is a one-to-one association
between motor states and source directions and they are indifferently denoted by $\{\xvect_n\}_{n=1}^N\in\mathcal{X}$. Note that the space $\mathcal{X}$ has a cylindrical \black{topology}.
The speaker is located in the far field of the head during the experiments ($>1.8$ meters), and Otani et al.\cite{OtaniHiraharaIse09}
showed that HRTFs mainly depend on the sound source direction while the distance
has \black{less} impact in that case. This is why sound source
locations are expressed with two angles in this work.

For each $\xvect_n\in\mathbb{R}^2$, two binaural recordings are performed: (i) white noise which can be used to estimate $\bar{\alphavect}_n$ and $\bar{\phivect}_n$, and (ii) a randomly selected utterance amongst 362 samples from the TIMIT dataset\cite{TIMIT}. These utterances are composed of 50$\%$ female, 50$\%$ male and they last 1 to 5 seconds. Sparse interaural spectrograms are computed from these recordings and are used to test our algorithms on natural sounds (Section~7).~
The resulting dataset is publicly available online and is referred to as the \textit{Computational Audio-Motor Integration through Learning} (CAMIL) dataset\fnm{b}\fnt{b}{http://perception.inrialpes.fr/$\sim$Deleforge/CAMIL\_Dataset/.}.

\section{The Manifold Structure of Interaural Cues}
\label{sec:manifold}

In this section, we analyze the intrinsic structure of the mean high- and low-ILD and -IPD vectors previously described. While these vectors live in a high-dimensional space, they might be parameterized by motor states and hence, could lie on a lower $L$-dimensional manifold, with $L=2$.
We propose to experimentally verify the existence of a Riemannian manifold structure\fnm{c}\fnt{c}{by definition, a Riemannian manifold is locally homeomorphic to a Euclidean space.} using non-linear dimensionality reduction, and examine whether \black{the} obtained representations are homeomorphic to the motor-state space.
Such a homeomorphism would allow us to validate or not the existence of a locally linear bijective mapping between motor states (or equivalently, source directions) and 
the interaural data gathered with our setup.

\subsection{Local Tangent Space Alignment}
If the interaural data lie in a linear subspace,
linear dimensionality reduction methods such as principal component analysis (PCA),
could be used. However, in the case of a non-linear subspace one should use a manifold learning technique, \textit{e.g.}, diffusion kernels as done in\cite{TalmonCohenGannot2011}.
We chose to use local tangent-space alignment (LTSA)\cite{ZhangZha04}
because it essentially relies on the assumption that the data are locally linear, which is our central hypothesis. 
LTSA starts by building a local neighborhood around each high-dimensional observation. Under the key assumption that each such neighborhood spans a linear space of low dimension $L$ corresponding to the dimensionality of the tangent space, \textit{i.e.}, a Riemannian manifold, PCA can be applied to each one of these neighborhoods thus yielding as many $L$-dimensional data representations as points in the dataset. 
Finally, a global map is built by optimal alignment of these local representations. This global alignment is done in the $L$-dimensional space by computing the $L$ largest eigenvalue-eigenvector pairs of a \textit{global alignment matrix} $\Bmat$ (see\cite{ZhangZha04} for details).
\begin{Figure}
    \centering
    \begin{tabular}{cccc}
      \includegraphics[width =0.2\linewidth,clip=,keepaspectratio]{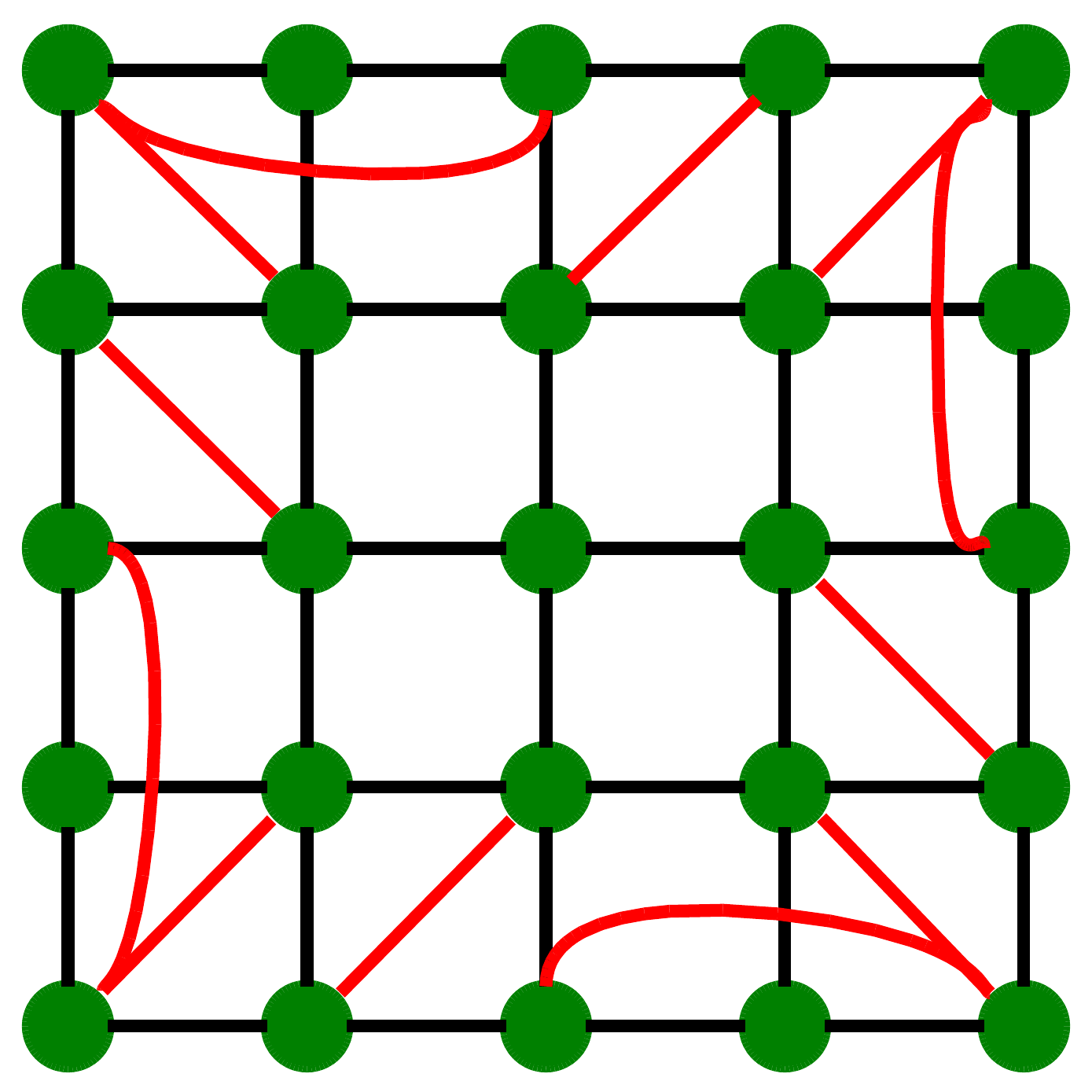}     &
      \includegraphics[width =0.2\linewidth,clip=,keepaspectratio]{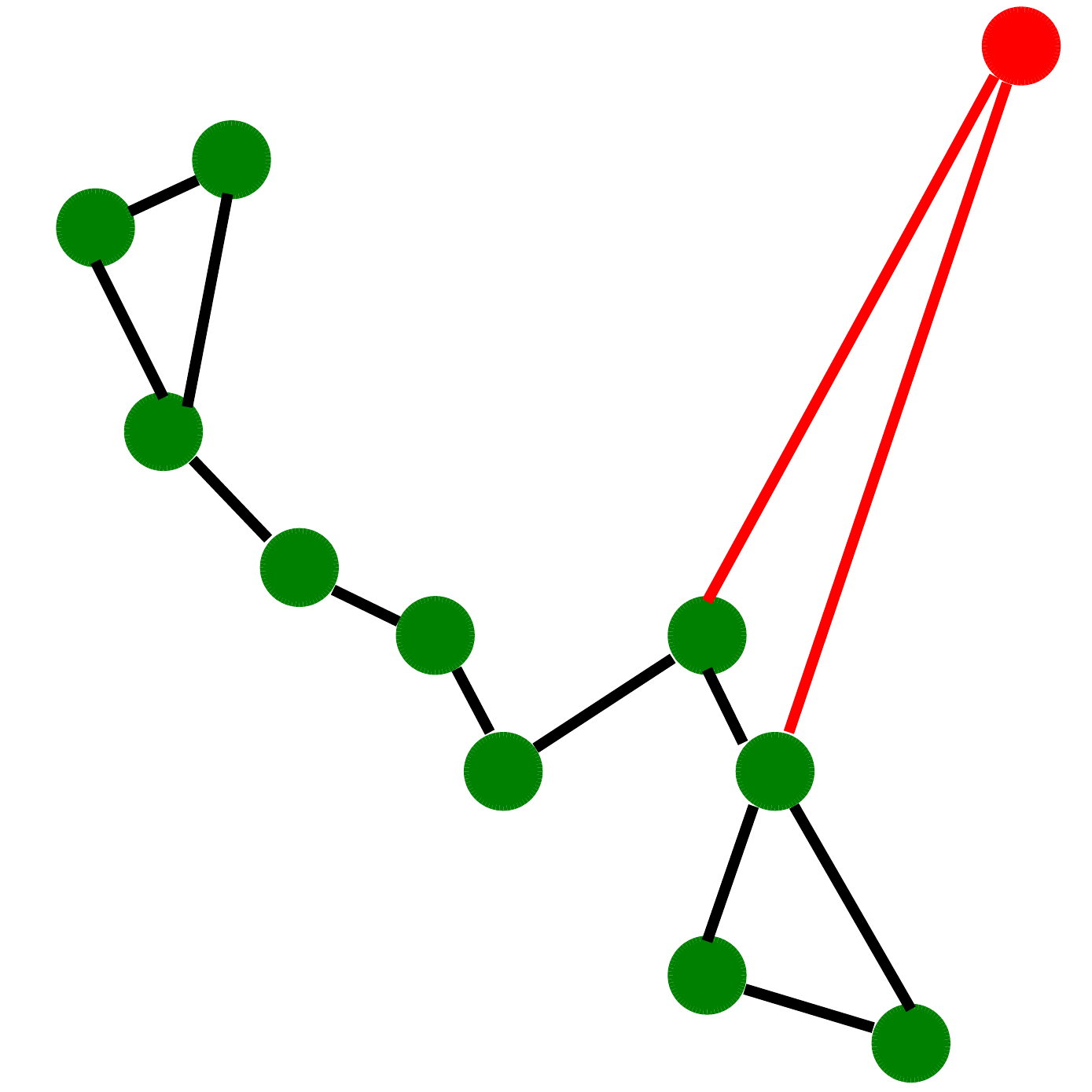}        &
      \includegraphics[width =0.2\linewidth,clip=,keepaspectratio]{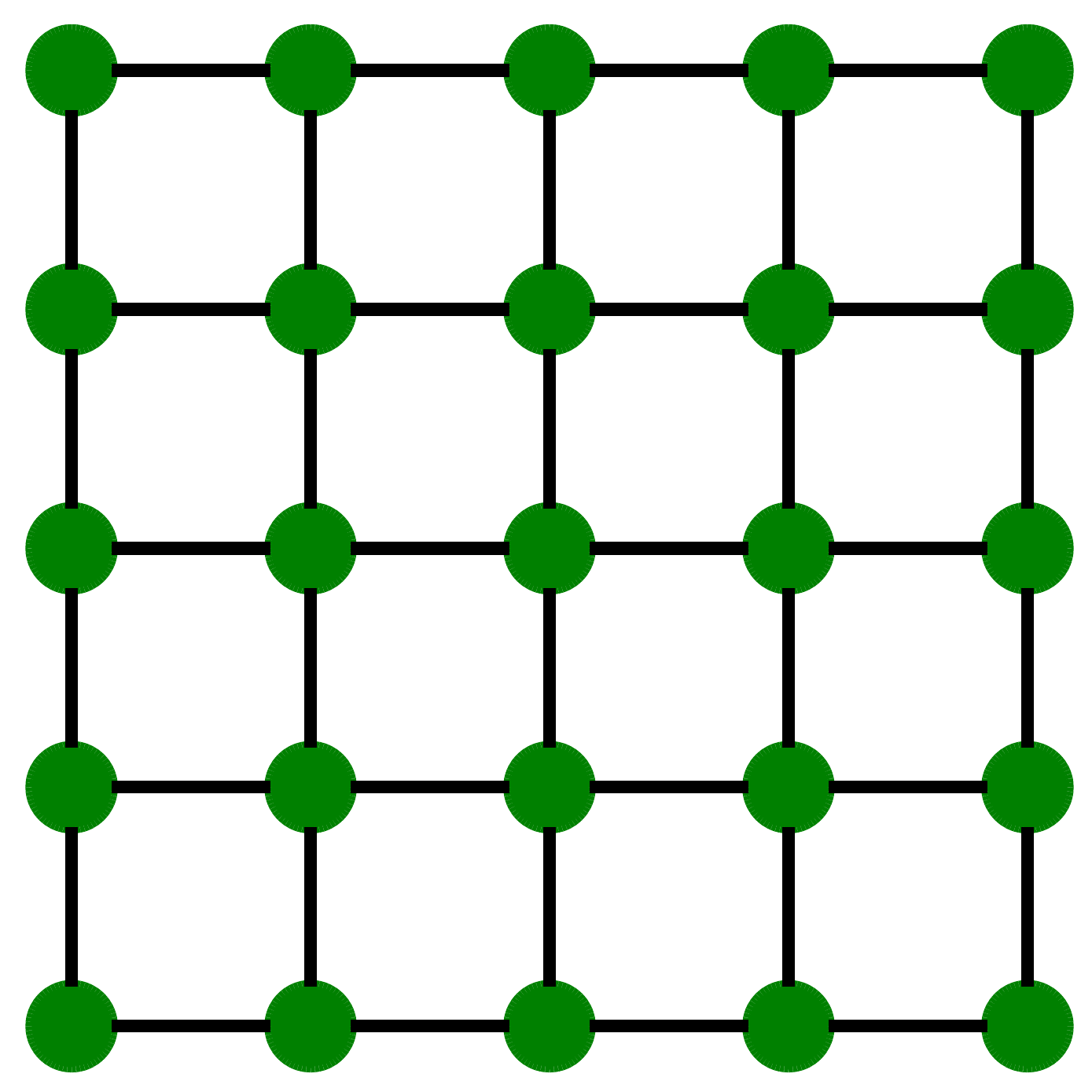} &
      \includegraphics[width =0.2\linewidth,clip=,keepaspectratio]{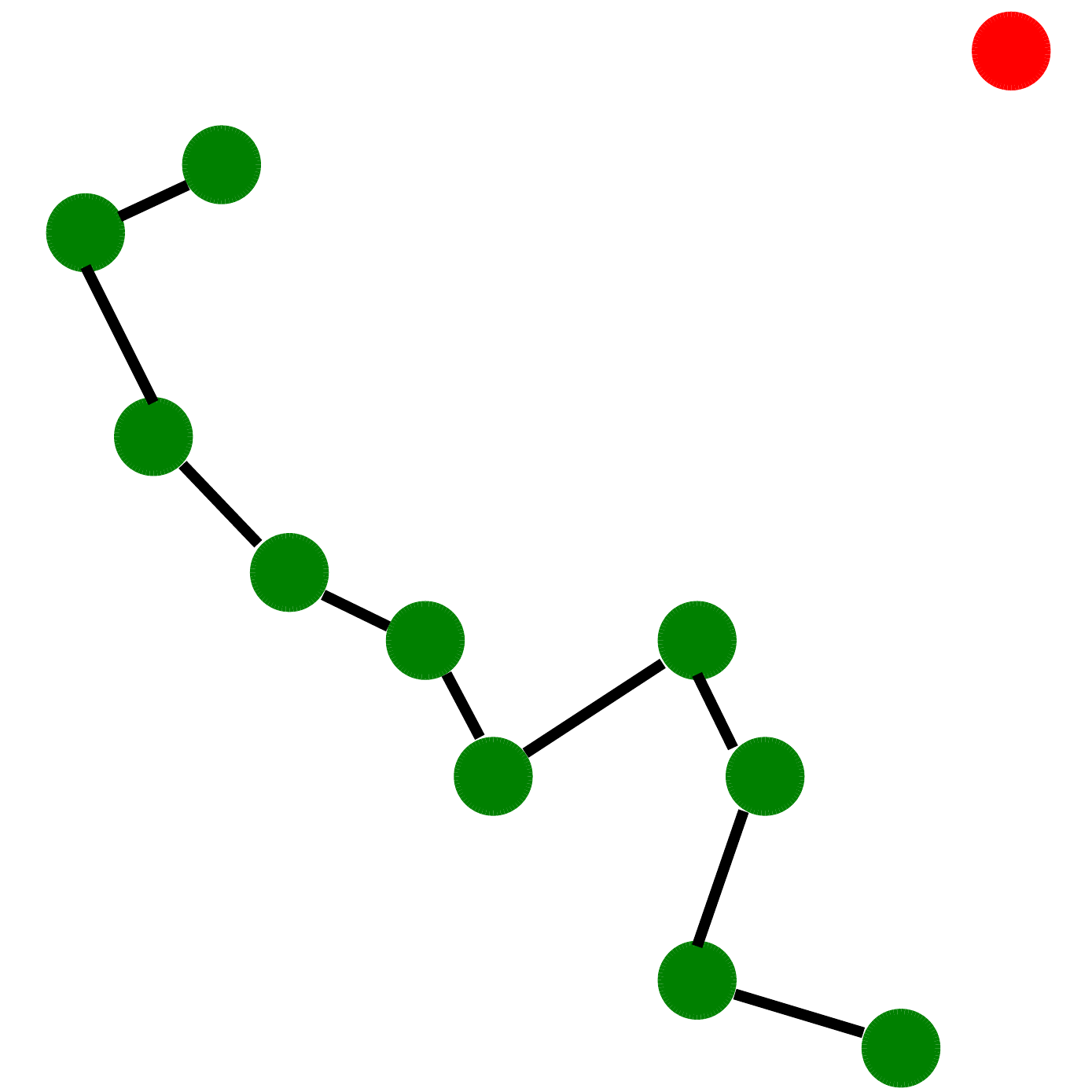}    \\
      (a) & (b) & (c) & (d) 
    \end{tabular}
    \captionof{figure}{\label{fig:neighborhoods}
      Differences between standard
      $k$NN (a,b) and symmetric $k$NN (c,d) on a grid
      of points with boundaries $(k=4)$ and
      in the presence of an outlier $(k=2)$.} 
\end{Figure}

Two extensions are proposed to adapt LTSA to our data. Firstly,
LTSA uses the $k$\textit{-nearest neighbors} ($k$NN) algorithm to determine neighboring
relationships between points, yielding neighborhoods of identical size $k$ over the data. This has the advantage of always providing graphs
with a single connected component, but it can easily lead to inappropriate edges
between points, especially at boundaries or in the presence of
outliers. A simple way to overcome these artifacts is to implement a
\textit{symmetric} version of $k$NN,
by considering that two
points are connected if and only if each of them belongs to the
neighborhood of the other one.
Comparisons between the outputs of standard and symmetric $k$NN are \black{shown} in Fig.~\ref{fig:neighborhoods}.
Although symmetric $k$NN solves connection issues at boundaries, it creates neighborhoods
of variable size, and in particular some points might get
disconnected from the graph. Nevertheless, it turns out that
ignoring such isolated points is an advantage since it may well be
viewed as a way to remove outliers from
the data. In our case the value of $k$ is set manually; in practice 
any value in the range $[15,25]$ yields satisfying results.

Secondly, LTSA is extended to represent manifolds \black{that} are
homeomorphic to the 2D surface of a cylinder. The best way
to visualize such a 2D curved surface is to represent it in the 3D
Euclidean space and to visualize the 3D points lying on that
surface. For this reason, we retain the $L+1=3$ largest 
eigenvalue-eigenvector pairs of the global alignment matrix $\Bmat$
such that the
extracted manifolds can be easily visualized.

\subsection{Binaural Manifold Visualization}
\begin{figure*}[t!]
   \begin{minipage}[c]{.64\linewidth}
      $\begin{array}{cc}
	\includegraphics[width = 0.50\linewidth,clip=,keepaspectratio]{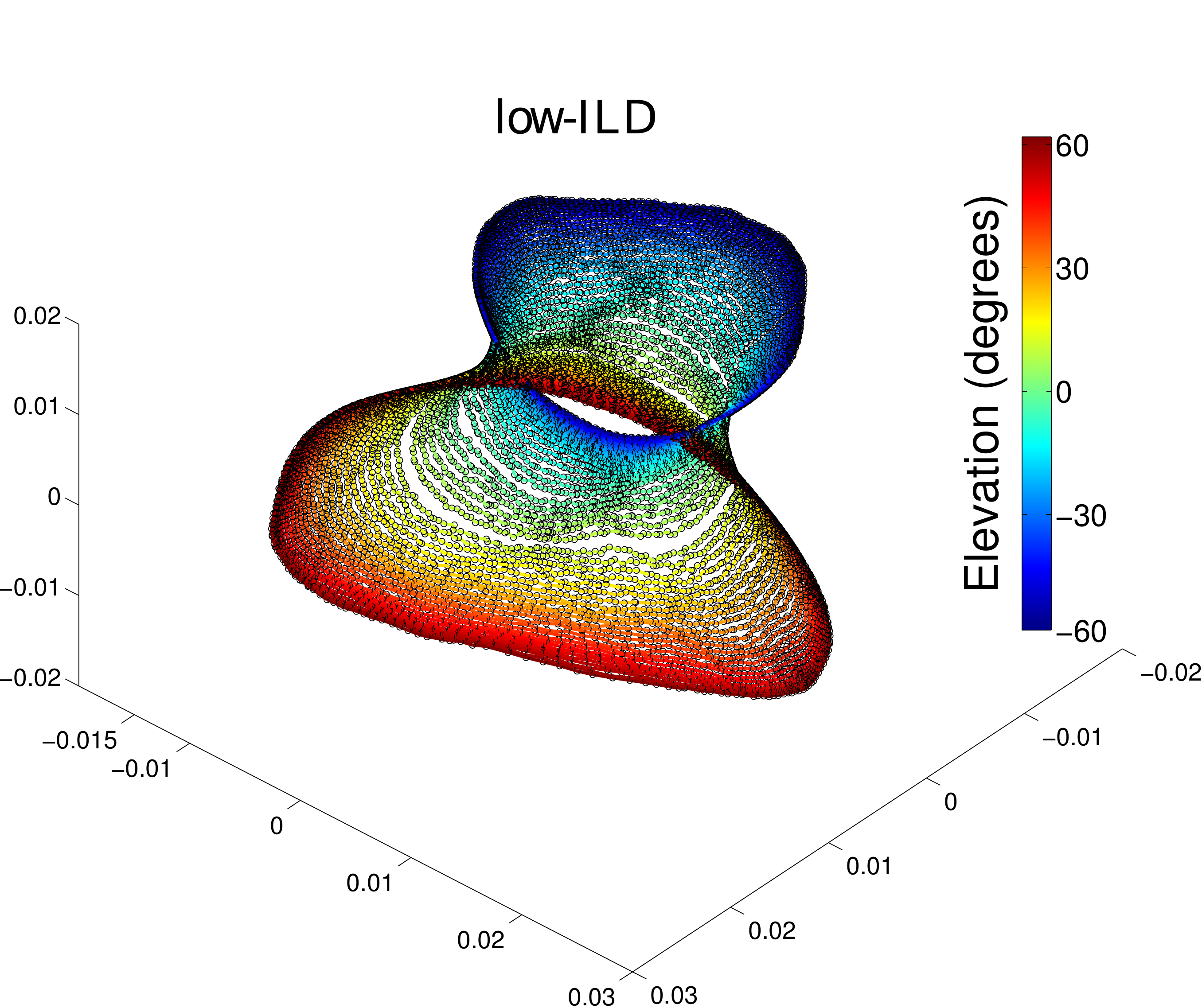} &
          \includegraphics[width = 0.50\linewidth,clip=,keepaspectratio]{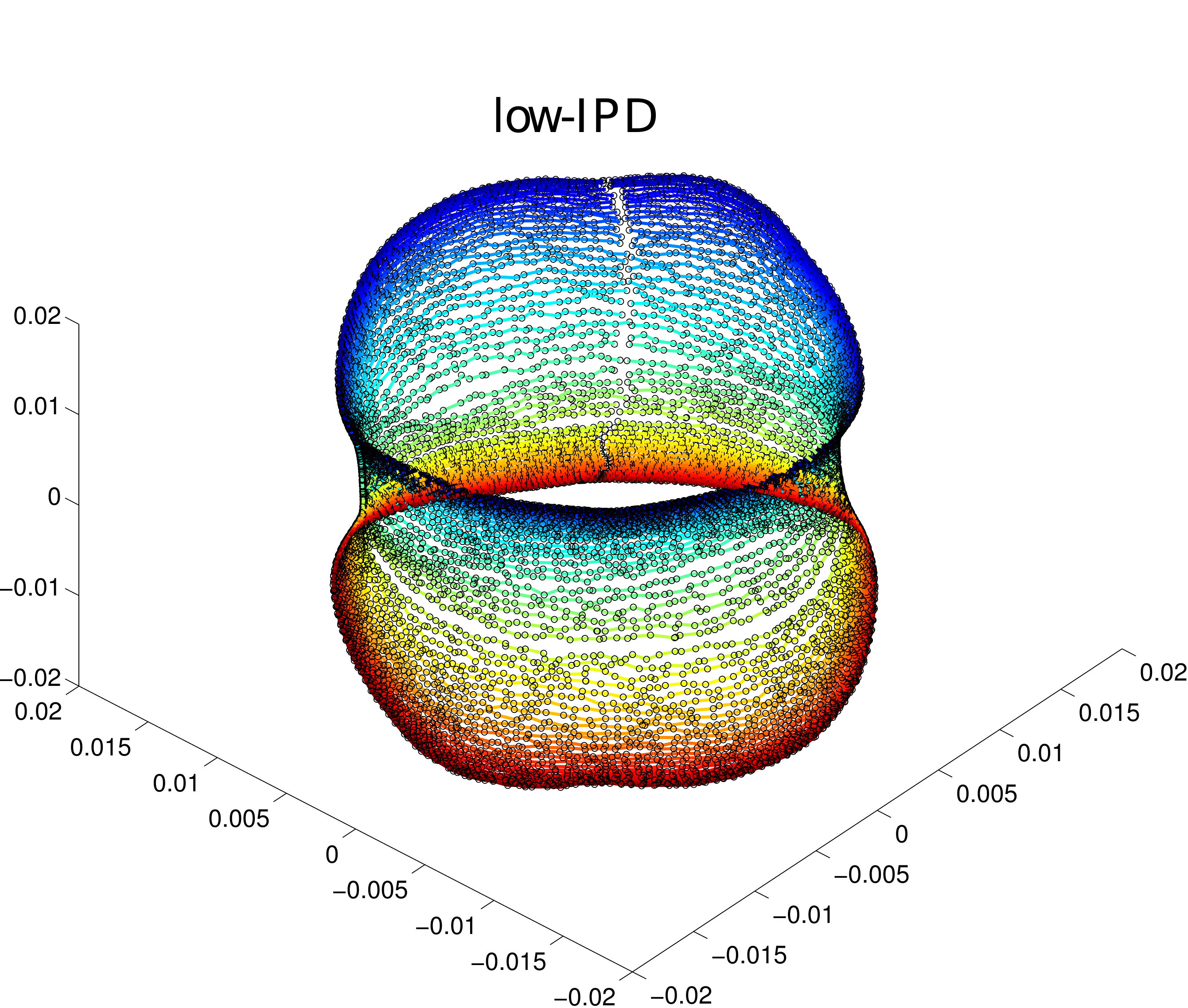} \\
        \includegraphics[width = 0.50\linewidth,clip=,keepaspectratio]{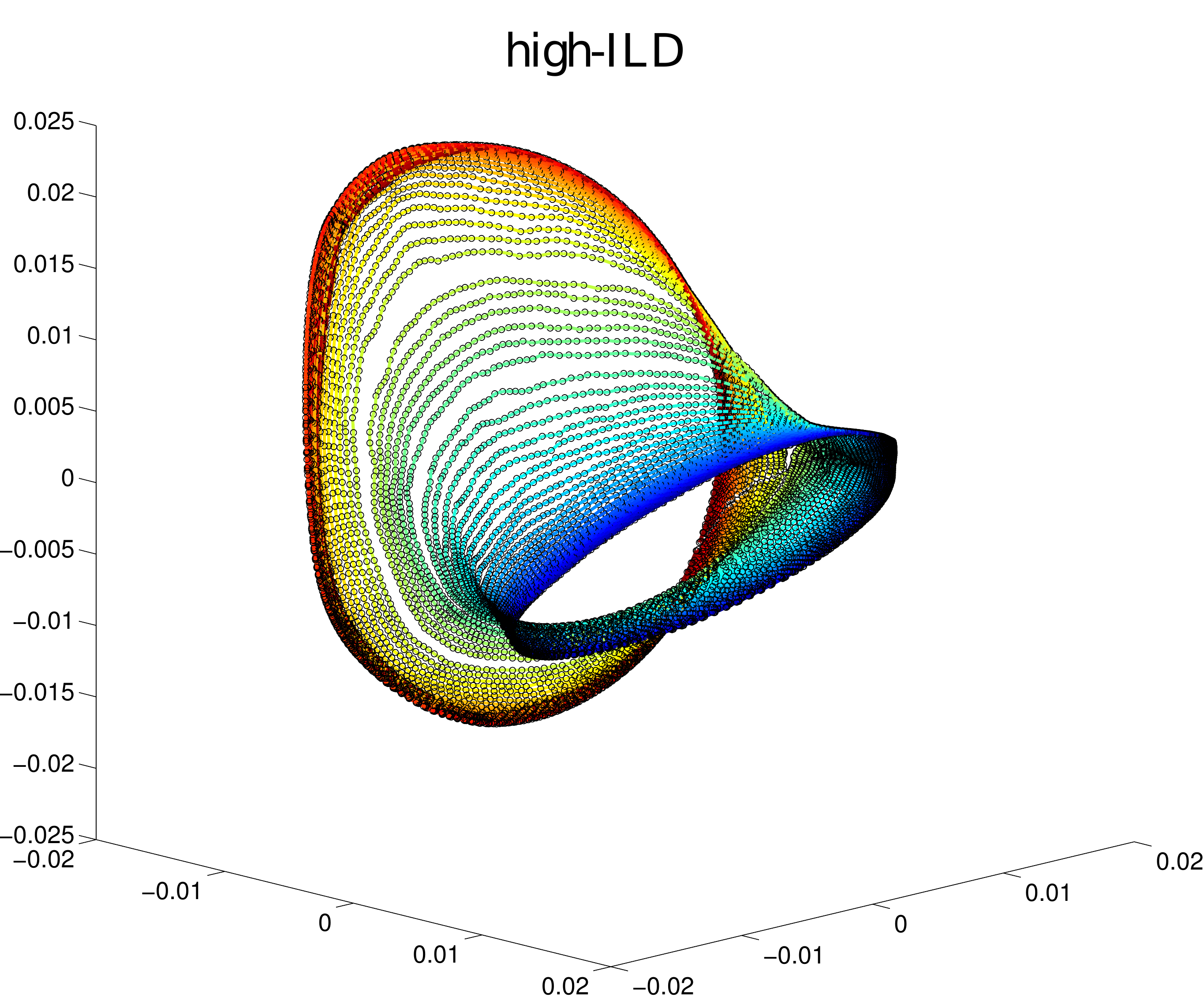} &
           \includegraphics[width = 0.50\linewidth,clip=,keepaspectratio]{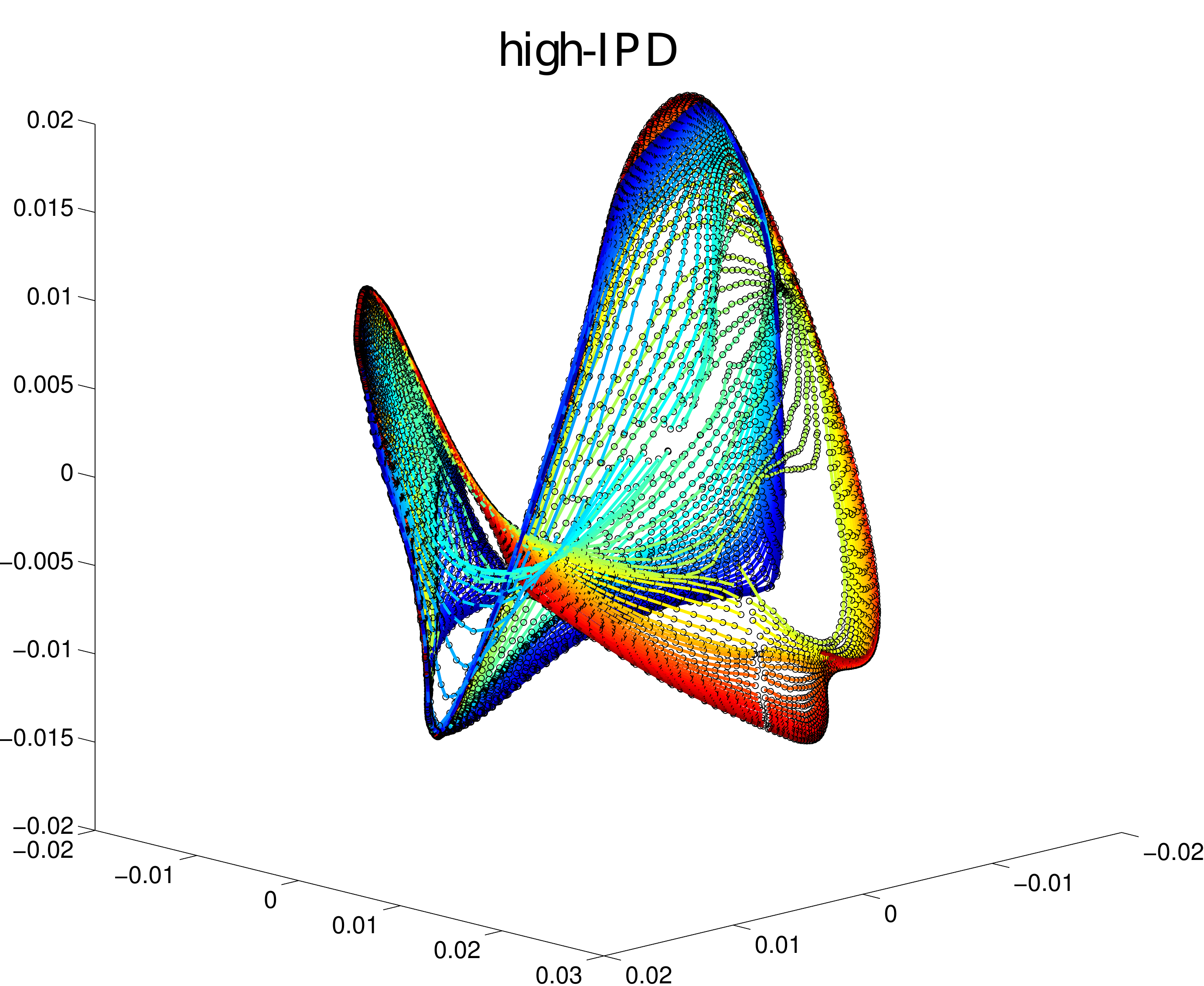}
      \end{array}$
      \caption{\small{\label{fig:emb} Low-dimensional representations of mean interaural vectors using non-linear dimensionality reduction (LTSA). For visualization purpose, points with the same ground truth elevation are linked with a colored line in azimuth order. \black{The three axes correspond to the first three eigenvectors of the global-alignment matrix.} Obtained point clouds are zero-centered and arbitrary scaled.}}
   \end{minipage} \hfill
   \begin{minipage}[c]{.32\linewidth}
      $\begin{array}{c}
      \includegraphics[width = 0.90\linewidth,clip=,keepaspectratio]{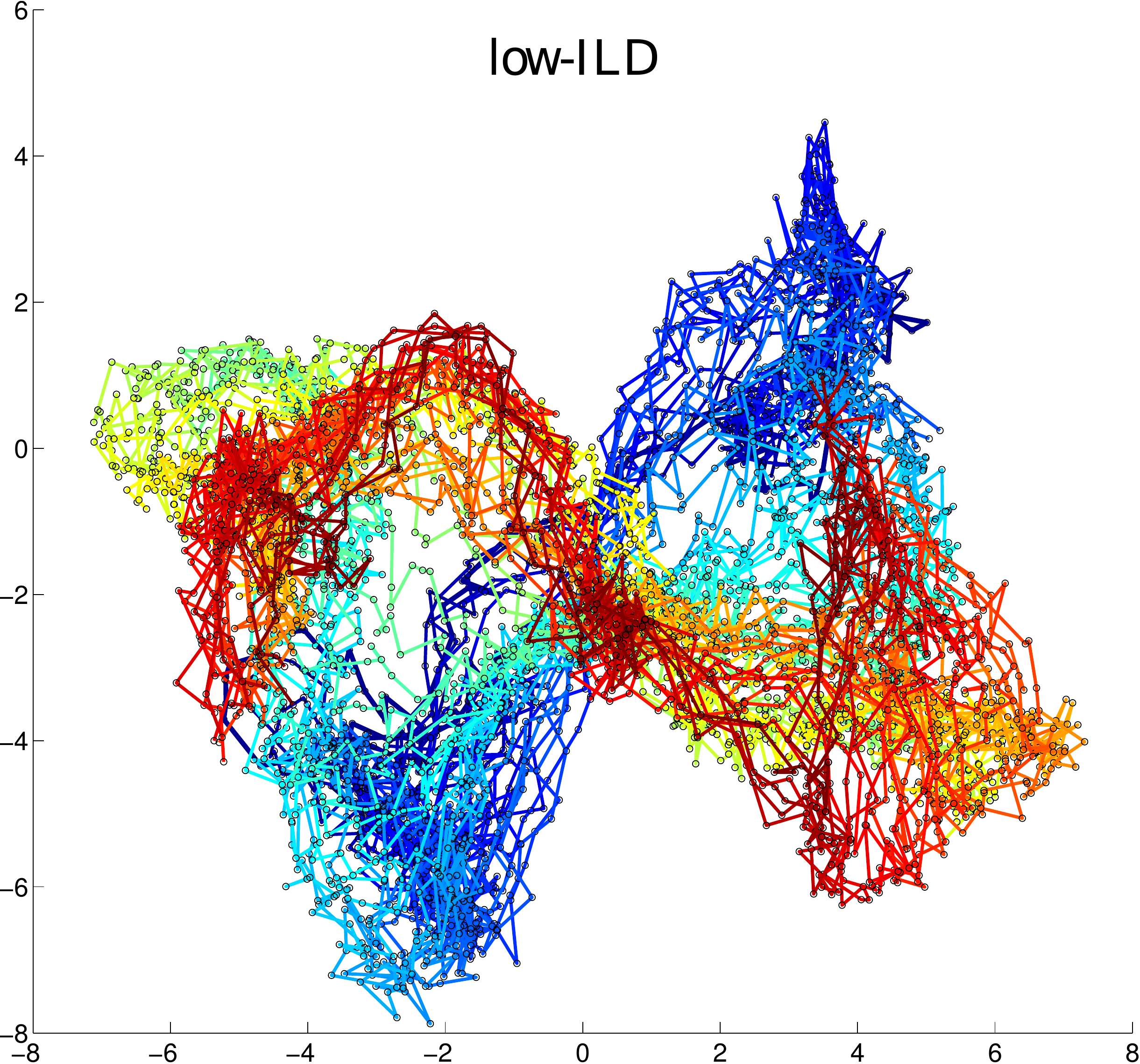} \\
      \includegraphics[width = 0.90\linewidth,clip=,keepaspectratio]{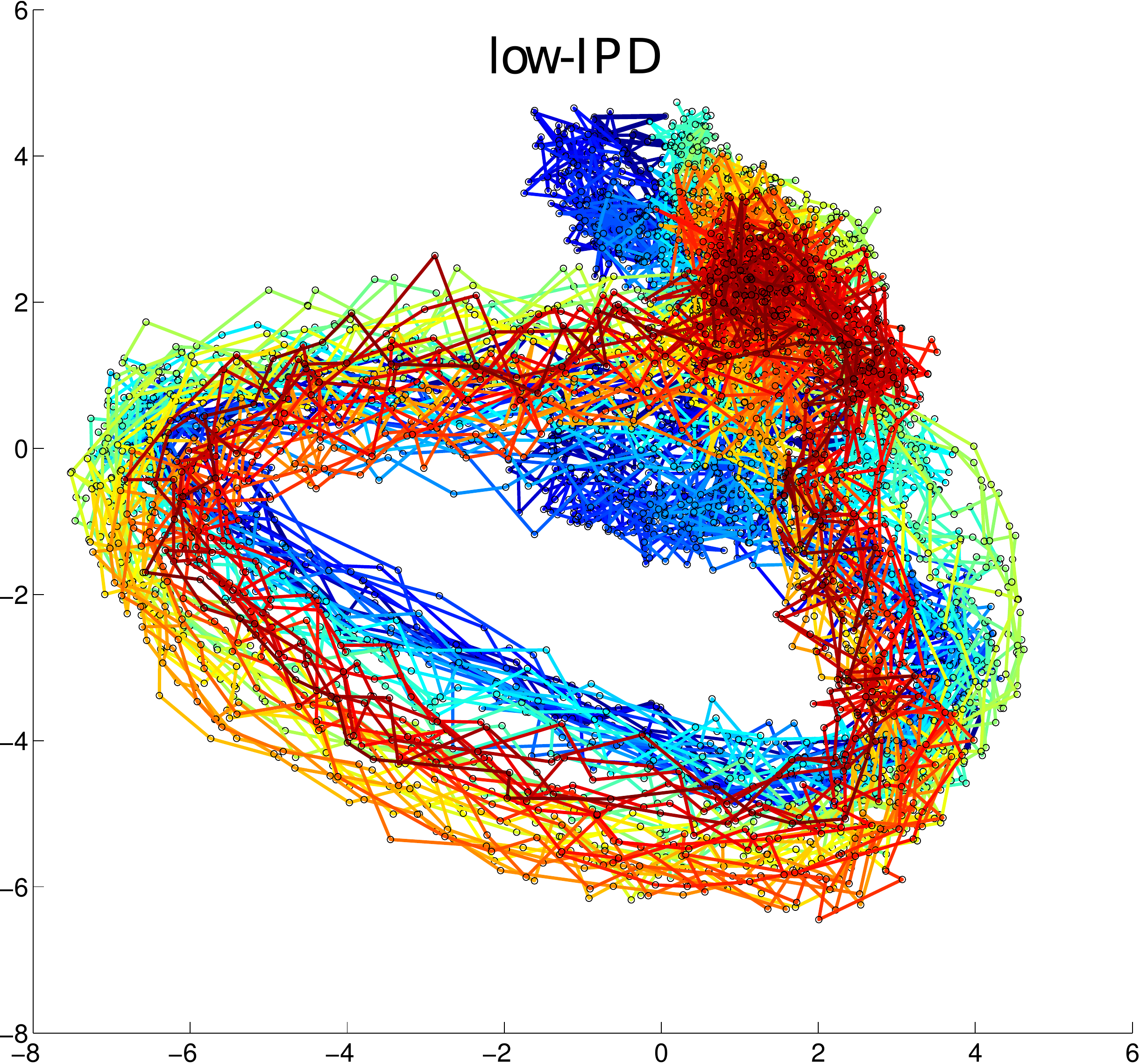} 
      \end{array}$
      \caption{\small{\label{fig:pca} Low-dimensional representations of mean interaural vectors using linear dimensionality reduction (PCA) (azimuths in $[-90^\circ,90^\circ]$). \black{The two axes correspond to the first two eigenvectors of the correlation matrix.}}}
   \end{minipage}
\end{figure*}

Low-dimensional representations obtained using LTSA are shown in Fig.~\ref{fig:emb}. Mean \textit{low}-ILD, \textit{low}-IPD, and \textit{high}-ILD surfaces are all smooth and homeomorphic to the motor-state space (a cylinder), thus confirming that these cues can be used for 2D binaural sound source localization based on locally linear mapping. However, this is not the case for the mean \textit{high}-IPD surface which features several distortions, elevation ambiguities, and crossings.  While these computational experiments confirm the duplex theory for IPD cues, they surprisingly suggest that ILD cues at low frequencies still contain rich 2D directional information. This was also hypothesized but not thoroughly analyzed in\cite{MandelWeissEllis10}. One can therefore concatenate full-spectrum ILD and low-frequency IPD vectors to form an observation space of dimension $730$, referred to as the \textit{ILPD space}. Similarly, one can define sparse \textit{ILPD spectrograms} for general sounds. These results experimentally \black{confirm} the existence of \textit{binaural manifolds}, \textit{i.e.}, a locally-linear, low-dimensional structure hidden behind the complexity of interaural spectral cues obtained from real world recordings in a reverberant room.

For comparison, Fig.~\ref{fig:pca} shows the result of applying PCA to $5,400$ mean \textit{low}-ILD vectors and \textit{low}-IPD vectors corresponding to frontal sources (azimuths in $[-90^\circ,90^\circ]$). The resulting representations are extremely distorted, due to the non-linear nature of binaural manifolds. This rules out the use of a linear regression method to estimate the interaural-to-localization mapping and justifies the development of an appropriate piecewise-linear mapping method, detailed in the next section.

\section{Probabilistic Acoustic Space Learning and Single Sound Source Localization}
\label{sec:ppam}
The manifold learning technique described above \black{allows retrieving}
intrinsic two-dimensional spatial coordinates from auditory cues and shows the existence of a smooth, locally-linear relationship between the two spaces. We now want to use these
results to address the more concrete problem of localizing a sound source. To do this, we need a technique that maps any high-dimensional interaural vector onto the $L=2$ dimensional space of sound source direction: azimuth and elevation. We will refer to the process of establishing such a mapping as \textit{acoustic space learning}. To be applicable to real world sound source localization, the mapping technique should feature a number of properties. First, it should deal with the sparsity of natural sounds mentioned in Section~2,~
and hence handle missing data. Second, it should deal with the high amount of noise and redundancy present in the interaural spectrograms of natural sounds, as opposed to the clean vectors obtained by averaging white noise interaural spectrograms during training (Section 2). Finally, it should allow further extension to the more complex case of mixtures of sound sources that will be addressed in Section~6.~
An attractive approach embracing all these properties is to use a Bayesian framework. We hence view the sound source localization problem as a probabilistic space mapping problem. This strongly contrasts with traditional approaches in sound source localization, which usually assume the mapping to be known, based on simplified sound propagation models. In the following section we present the proposed model, which may be viewed as a variant of the mixture of local experts model\cite{XuJordanHinton95} with a geometrical interpretation.

\subsection{Probabilistic Piecewise Affine Mapping}
\label{subsec:ppam}
\vspace{-5mm}
\paragraph{Notations}
Capital letters indicate random variables, and
lower case their realizations. \black{We both use the following equivalent notations:} $p(V=v)$ and
$p(v)$ for the probability or density of variable $V$ at value $v$. Subscripts $n\in\{1\dots N\}$, $d\in\{1\dots D\}$, $k\in\{1\dots K\}$, $t\in\{1\dots T\}$ and $m\in\{1\dots M\}$ are
respectively indexes over training observations, interaural vector components, affine transformations, spectrogram time frames and sources.

As mentioned in Section~3,~
the complete space of sound source positions (or motor states) used in the last sections has a cylindrical topology. However, to make the inference of the mapping model described below analytically tractable, it is preferable for the low-dimensional space to have a linear (Euclidean) topology. Indeed, this notably allows standard Gaussian distributions \black{to be defined} over that space which are easy to deal with in practice. To do so, we will simply use a subset of the complete space of sound source positions, corresponding to azimuths between $-160^\circ$ and $160^\circ$ degrees. This subset will be used throughout the article. To generalize the present work to a cylindrical space of source positions, one could imagine learning two different mapping models: one for frontal sources and one for rearward sources. Localization could then be done based on a mixture of these two models.

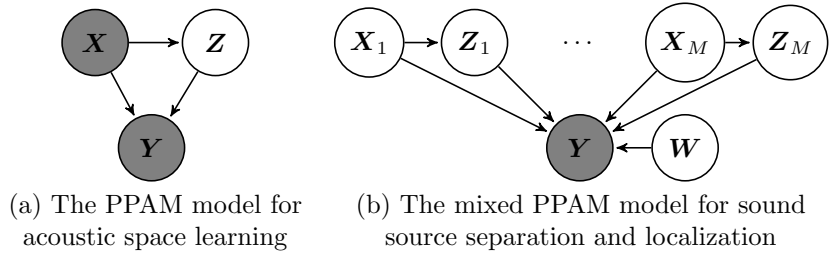
\begin{figure*}[t!]
\centering
\begin{tabular}{cc}
\begin{tikzpicture}[->,>=stealth',shorten >=1pt,auto,node distance=1.6cm,
                    semithick]
  \tikzstyle{every state}=[circle]
  \node[state,fill=gray] (X) {$\Xvect$};
  \node[state] (Z) [right of=X] {$\Zvect$};
  \node[state,fill=gray] (Y) [right=0.75cm,below=0.95cm] at (X)  {$\Yvect$};
  \path (X) edge (Z)
    (X) edge (Y)
    (Z) edge (Y);
\end{tikzpicture} &
\begin{tikzpicture}[->,>=stealth',shorten >=1pt,auto,node distance=1.4cm,
                    semithick]
  \tikzstyle{every state}=[circle]
  \node[state] (X1) {$\Xvect_1$};
  \node[state] (Z1) [right of=X1] {$\Zvect_1$};
  \node[draw=white] (DOTS) [right of=Z1] {$\dots$};
  \node[state] (XM) [right of=DOTS] {$\Xvect_M$};
  \node[state] (ZM) [right of=XM] {$\Zvect_M$};
  \node[state,fill=gray] (Y) [below of=DOTS] {$\Yvect$};
  \node[state] (W) [right of=Y] {$\Wvect$};
  \path (X1) edge (Z1)
    (X1) edge (Y)
    (Z1) edge (Y)
    (XM) edge (ZM)
    (XM) edge (Y)
    (ZM) edge (Y)
    (W)  edge (Y);
\end{tikzpicture} \\
 (a) The PPAM model for & (b) The mixed PPAM model for sound\\
 acoustic space learning  &  source separation and localization
\end{tabular}
\caption{\small{Graphical models of PPAM and mixed PPAM. White means unobserved, gray means observed.}}
\label{fig:graphical_model}
\end{figure*}

Let $\mathcal {X}\subset\mathbb{R}^L$ denote the low-dimensional space of source positions and $\mathbb{R}^D$ the high-dimensional observation space, \textit{i.e.}, the space of interaural cues. The computational experiments of Section~4~
suggest that there exists a smooth, locally linear bijection $g:\mathcal{X}\rightarrow\mathcal{Y}\subset\mathbb{R}^D$ such that the set $\mathcal{Y}=\{g(\xvect), \xvect\in\mathcal{X}\}$ forms an $L-$dimensional manifold embedded in $\mathbb{R}^D$.
Based on this assumption, the proposed idea is to compute a piecewise-affine probabilistic approximation of $g$ from a training data set $\{(\xvect_n,\yvect_n)\}_{n=1}^N\subset \mathcal{X}\times \mathcal{Y}$ and to estimate the inverse of $g$ using a Bayesian formulation.
The local linearity of $g$ suggests that each point $\yvect_n$ is the image of a point $\xvect_n\in\mathcal{R}_k \subset \mathcal{X}$ by an affine transformation $\tau_k$, plus an error term. Assuming that there is a finite number $K$ of such affine transformations $\tau_k$ and an equal number of associated regions $\mathcal{R}_k$, we obtain a piecewise-affine approximation of $g$. An assignment variable  $z_n \in \{1\dots K\}$ is associated with each training pair $(\xvect_n,\yvect_n)$ such that $z_{n}=k$ if $\yvect_n$ is the image of $\xvect_n\in\mathcal{R}_k$ by
$\tau_k$. 
This allows us to write ($\mathbb{I}_{\{z_n=k\}}=1$ if $z_n=k$,
and 0 otherwise):
\begin{equation}
\label{eq:point_k}
  \yvect_n=\textstyle{\sum_{k=1}^K }\mathbb{I}_{\{z_n=k\}} (\Avect_k \xvect_n + \bvect_k) + \evect_n
\end{equation}
where the $D\times L$ matrix $\Amat_k$ and the vector $\bvect_k\in\mathbb{R}^D$ define the transformation $\tau_k$, and $\evect_n\in\mathbb{R}^D$ is an error term capturing both the observation noise and the reconstruction error of affine transformations. If we make the assumption that the error terms $\evect_n$ do not depend on $\xvect_n$, $\yvect_n$ or $\zvect_n$, and are \black{independent} identically distributed realizations of a Gaussian variable with $\zerovect$ mean and diagonal covariance matrix $\Sigmamat=\operatorname{diag}(\sigma^2_{1:D})$, we obtain:
\begin{equation}
\label{eq:learning_model_y}
 p(\yvect_n|\xvect_n,Z_{n}=k;\thetavect) =\mathcal{N}(\yvect_n ; \Amat_k \xvect_n + \bvect_k,\Sigmamat)
\end{equation}
where $\thetavect$ designates all the model parameters (see (\ref{eq:learning_model_Theta})). To make the affine transformations local, we set $z_n$ to the realization of a hidden multinomial random variable $Z_n$ conditioned by $\xvect_n$:
\begin{equation}
 \label{eq:learning_model_z}
 p(Z_{n}=k|\xvect_n ; \thetavect) = \frac{\pi_k\mathcal{N}(\xvect_n ; \cvect_k,\Gammamat_k)}{\sum_{k=1}^K\pi_k\mathcal{N}(\xvect_n ; \cvect_k,\Gammamat_k)}, \\
\end{equation}
where $\cvect_k\in\mathbb{R}^L$, $\Gammamat_k\in\mathbb{R}^{L\times L}$ and $\sum_k\pi_k=1$. We can give a geometrical interpretation of this distribution by adding the following \textit{volume equality} constraints to the model:
\begin{equation}
\label{eq:volume-constraints}
 |\Gammamat_1|= \dots = |\Gammamat_K| \hspace{0.1cm} \textrm{and} \hspace{0.2cm}  \pi_1=  \dots =\pi_K = 1/K
\end{equation}
One can verify that under these constraints, the set of $K$ regions of $\mathcal{X}$ maximizing (\ref{eq:learning_model_z}) for each $k$ defines a Voronoi diagram of centroids $\{\cvect_k\}_{k=1}^K$, where the Mahalanobis distance $||.||_{\Gammamat_k}$ is used instead of the Euclidean one. This corresponds to a compact probabilistic way of representing a general partitioning of the low-dimensional space into convex regions of equal volume. Extensive tests on simulated and audio data showed that these constraints yield lower reconstruction errors, on top of providing a meaningful interpretation of (\ref{eq:learning_model_z}). To make our generative model complete, we define the following Gaussian mixture prior on $\Xvect$:
\begin{equation}
 \label{eq:learning_model_x}
 p(\Xvect_n=\xvect_n ; \thetavect)=\textstyle\sum_{k=1}^K\pi_k\mathcal{N}(\xvect_n ; \cvect_k,\Gammamat_k)
\end{equation}
which allows (\ref{eq:learning_model_z}) and
(\ref{eq:learning_model_x}) \black{to be rewritten} in a simpler form: 
\begin{equation}
\label{priorXZsimple}
p(\Xvect_n=\xvect_n, Z_n=k ; \thetavect)= \pi_k\mathcal{N}(\xvect_n ;\cvect_k,\Gammamat_k).
\end{equation}
The graphical model of PPAM is given in Fig.~\ref{fig:graphical_model}(a) and Fig.~\ref{fig:ppam} shows a partitioning example obtained with PPAM using a toy data set.  
\begin{Figure}
\centering
\includegraphics[width = \linewidth,clip=,keepaspectratio]{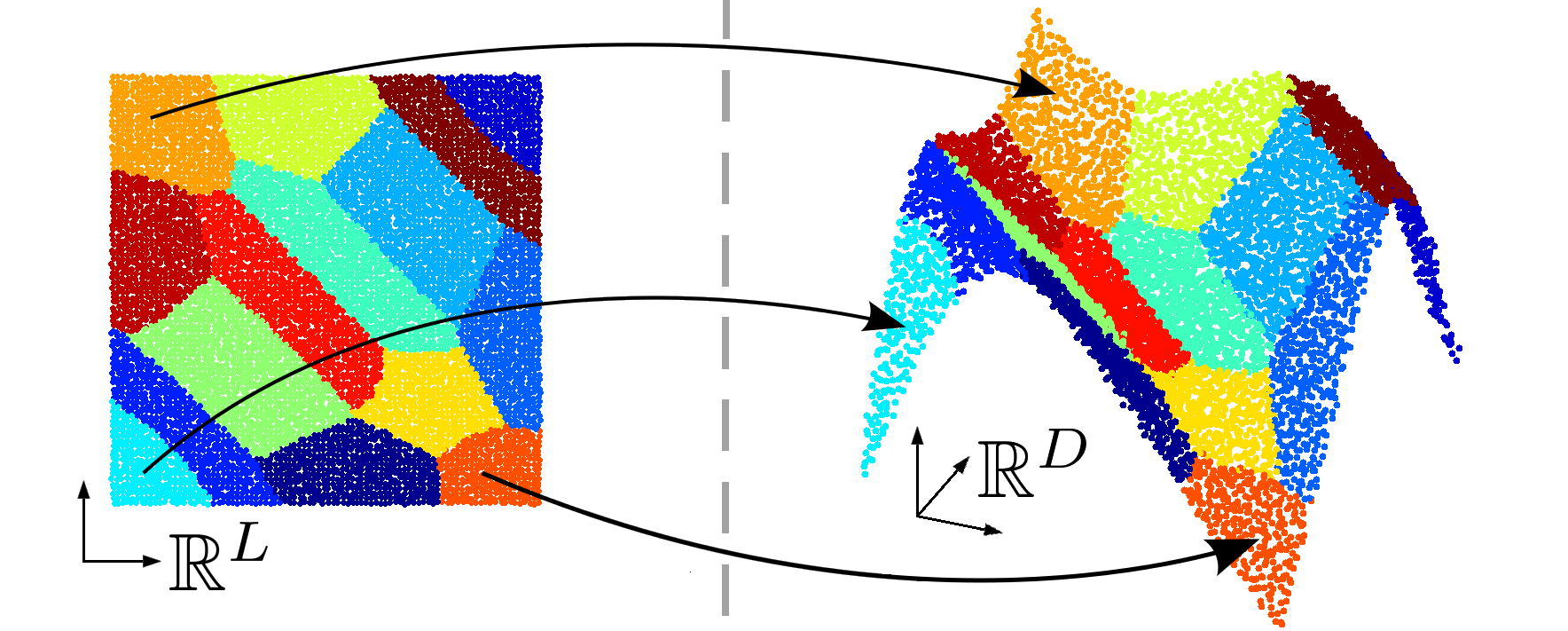}
\captionof{figure}{{\label{fig:ppam} Space partitioning and locally affine mapping on a toy data set ($N=9600,K=15,L=2,D=3$). Colors encode regions in $\mathbb{R}^D$ maximizing (\ref{eq:learning_model_z}). Observe how these regions (associated with affine transformations) are adjusted to the geometry of the observation manifold.}}
\end{Figure}

The inference of PPAM can be \black{achieved} with a closed-form and efficient EM algorithm maximizing the observed-data log-likelihood $\log p(\Xmat,\Ymat ; \thetavect)$ with respect to the model parameters:
\begin{equation}
 \label{eq:learning_model_Theta}
 \thetavect = \left\{\{\Gammamat_k,\cvect_k,\Amat_k,\bvect_k\}_{k=1}^K,\Sigmamat\right\}.
\end{equation}
\black{The EM algorithm is made of two steps: E (Expectation) and M (Maximization). In the E-step posteriors
$r_{kn}^{(i)} = p(Z_{n}=k|\xvect_n,\yvect_n;\thetavect^{(i-1)})$ at iteration $i$
are computed from (\ref{eq:learning_model_y}), (\ref{eq:learning_model_z}) and Bayes inversion. In the M-step we maximize the expected complete-data log-likelihood $\mathbb{E}_{(\Zvect|\Xvect,\Yvect,\thetavect^{(i)})}[\log p(\Xvect,\Yvect,\Zvect|\thetavect)]$ with respect to parameters $\thetavect$.} We obtain the following closed-form expressions for the parameters updates under the volume equality constraints (\ref{eq:volume-constraints}):
\begin{align}
 \label{eq:ppam_mstep}
 \cvect_k^{(i)} &= \sum_{n=1}^N\frac{r^{(i)}_{kn}}{\bar{r}^{(i)}_{k}}\xvect_n, \hspace{0.1cm} \Gammamat_k^{(i)} = \frac{\Smat^{(i)}_k}{|\Smat^{(i)}_k|^{\frac{1}{L}}}\sum_{j=1}^K\frac{\bar{r}^{(i)}_j}{N}|\Smat^{(i)}_j|^{\frac{1}{L}} \nonumber \\
 \Amat_k^{(i)} &= \overline{\Ymat}^{(i)}_k{\overline{\Xmat}^{(i)\dagger}_k}, \hspace{0.2cm} \bvect_k^{(i)} = \sum_{n=1}^N\frac{r^{(i)}_{kn}}{\bar{r}^{(i)}_{k}}(\yvect_n-\Amat^{(i)}_k\xvect_n), \nonumber \\
 {\sigma^{2}}^{(i)}_d &= \frac{1}{K}\sum_{k=1}^K\sum_{n=1}^N\frac{r^{(i)}_{kn}}{\bar{r}^{(i)}_{k}}(y_{dn}-{\avect_{dk}^{(i)\top}}\xvect_n-b_{dk}^{(i)})^2
\end{align}
where $^{\dagger}$ is the Moore-Penrose pseudo inverse operator, $(.,.)$ denotes horizontal concatenation and:
\begin{align*}
 \Smat^{(i)}_k&={\textstyle \sum_{n=1}^N}r^{(i)}_{kn}/\bar{r}^{(i)}_{k}(\xvect_n-\cvect^{(i)}_k)(\xvect_n-\cvect^{(i)}_k)\tp\\
 \bar{r}^{(i)}_{k}&={\textstyle \sum_{k=1}^K}r^{(i)}_{kn}, \hspace{0.2cm} \Amat^{(i)}_k=(\avect_{1k}^{(i)}, \dots, \avect_{Dk}^{(i)})\tp\\
 \overline{\Xmat}^{(i)}_k &= (r_{k1}^{(i)\frac{1}{2}}(\xvect_1-\bar{\xvect}^{(i)}_k) \dots r_{kN}^{(i)\frac{1}{2}}(\xvect_N-\bar{\xvect}^{(i)}_k))\\
 \overline{\Ymat}^{(i)}_k &= (r_{k1}^{(i)\frac{1}{2}}(\yvect_1-\bar{\yvect}^{(i)}_k) \dots r_{kN}^{(i)\frac{1}{2}}(\yvect_N-\bar{\yvect}^{(i)}_k))\\
 \bar{\xvect}^{(i)}_k &= {\textstyle \sum_{n=1}^N}r^{(i)}_{kn}/\bar{r}^{(i)}_{k}\xvect_n, \;\bar{\yvect}^{(i)}_k = {\textstyle \sum_{n=1}^N}r^{(i)}_{kn}/\bar{r}^{(i)}_{k}\yvect_n.
\end{align*}
Initial posteriors $r_{kn}^{(0)}$ can be obtained either by estimating a $K$-GMM solely on $\Xvect$ or on joint data $[\Xvect;\Yvect]$ ($[.;.]$ denotes vertical concatenation) and then go on with the M-step (\ref{eq:ppam_mstep}). The latter strategy generally provides a better initialization and hence a faster convergence, at the cost of being more computationally demanding.

Given a set of parameters $\thetavect$, a mapping from a
test position $\xvect_t\in\mathbb{R}^L$ to its corresponding interaural cue $\yvect_t\in\mathbb{R}^D$ is obtained using the
\textit{forward conditional density} of PPAM, \textit{i.e.}, $p(\yvect_t|\xvect_t;\thetavect)=$
\begin{equation}
 \label{eq:JGMM_forward_map}
\frac{ \sum_{k=1}^K \pi_k\mathcal{N}(\xvect_t ; \cvect_k,\Gammamat_k) \mathcal{N}(\yvect_t;\Amat_k\xvect_t+\bvect_k,\Sigmamat)}{\sum_{j=1}^K\pi_j\mathcal{N}(\xvect_t ; \cvect_j,\Gammamat_j)}
\end{equation}
while a mapping from an interaural cue to its corresponding source position is obtained using the \textit{inverse conditional density}, \textit{i.e.}, $p(\xvect_t|\yvect_t;\thetavect) =$
\begin{equation}
 \label{eq:JGMM_inverse_map}
\frac{\sum_{k=1}^K \pi_k\mathcal{N}(\yvect_t ; \cvect_k^*,\Gammamat_k^*) \mathcal{N}(\xvect_t;\Amat^*_k\yvect_t+\bvect^*_k,\Sigmamat_k^*)}{\sum_{j=1}^K\pi_j\mathcal{N}(\yvect_t ; \cvect_j^*,\Gammamat^*_j)}
 \end{equation}
where $\cvect_k^{\ast}=\Amat_k\cvect_k+\bvect_k$,
$\Gammamat_k^{\ast}=\Sigmamat+\Amat_k\Gammamat_k\Amat_k\tp$,
$\Amat^{\ast}_k = \Sigmamat^{\ast}\Amat_k\tp\Sigmamat^{-1}$,
$\bvect^{\ast}_k =
\Sigmamat_k^*(\Gammamat_k^{-1}\cvect_k-\Amat_k\tp\Sigmamat^{-1}\bvect_k)$,
and $\Sigmamat_k^{\ast} =
(\Gammamat_k^{-1}+\Amat_k\tp\Sigmamat^{-1}\Amat_k)^{-1}$. Note
that both
(\ref{eq:JGMM_forward_map}) and (\ref{eq:JGMM_inverse_map}) take
the form of a Gaussian mixture distribution in one space given an observation in the other space.
These Gaussian mixtures are parameterized in two different ways by the
observed data and PPAM's parameters.
One can use their expectations $\mathbb{E}[\Yvect_t|\xvect_t;\thetavect]$ and $\mathbb{E}[\Xvect_t|\yvect_t;\thetavect]$ to obtain
\textit{forward} and \textit{inverse} mapping functions.
The idea of learning the mapping from the low-dimensional space of source position to the high-dimensional space of auditory observations and then \black{inverting} the mapping for sound source localization using (\ref{eq:JGMM_inverse_map}) is crucial. It implies that the number of parameters $\thetavect$ to estimate is $K(D(L+2)+L+L^2+1)$ while it would be $K(L(D+2)+D+D^2+1)$ if the mapping was learned the other way around. The latter number is 130 times the former for $D=512$ and $L=2$, making the EM estimation impossible in practice due to over-parameterization.

\vspace{-4mm}
\subsection{Localization From Sparse Spectrograms}
\label{subsec:single_source}
So far, we have considered mapping from a vector space to another vector space. However, as detailed in Section~2,~
spectrograms of natural sounds are time series of possibly noisy sparse vectors.
The Bayesian framework presented above allows to deal with this situation straightforwardly. Let $\widetilde{\thetavect}$ denote a set of parameters estimated with PPAM and let $\Yvect_{\chivect}=\{y_{dt};\chi_{dt}\}_{t,d=1}^{T,D}$ be an observed sparse interaural spectrogram ($\chi_{dt}$ denotes missing values as defined in Section~2.2).
Note that $d$ may represent the index of an ILD or an IPD value at a given frequency. If we suppose that all the observations are assigned to the same source position $\xvect$ and transformation $z$, it follows from the model (\ref{eq:learning_model_y}),  (\ref{eq:learning_model_z}), (\ref{eq:learning_model_x}) that the posterior distribution $p(\xvect|\Yvect_{\chivect};\widetilde{\thetavect})$ is a GMM $\sum_{k=1}^K\rho_k\mathcal{N}(\xvect;\mvect_k,\Vmat_k)$ in $\mathbb{R}^L$ with parameters:
\begin{align}
\label{eq:spectrogram_gmm_m}
\mvect_k &= \Vmat_k\biggl(\widetilde{\Gammamat}_k^{-1}\widetilde{\cvect}_k+{\sum_{d,t=1}^{D,T}}\frac{\chi_{dt}}{\widetilde{\sigma}^2_d}\widetilde{\avect}_{dk}(y_{dt}-\widetilde{b}_{dk})\biggr), \\
\label{eq:spectrogram_gmm_V}
\Vmat_k &= \biggl(\widetilde{\Gammamat}_k^{-1}+{\textstyle\sum_{d,t=1}^{D,T}}\frac{\chi_{dt}}{\widetilde{\sigma}^2_d}\widetilde{\avect}_{dk}\widetilde{\avect}_{dk}\tp\biggr)^{-1}\textrm{and}\hspace{0.2cm} 
\end{align}
\begin{align}
\label{eq:spectrogram_gmm_rho}
\rho_k &\propto \frac{|\Vmat_k|^{\frac{1}{2}}}{|\widetilde{\Gammamat}_k|^{\frac{1}{2}}}\exp\biggl(-\frac{1}{2}\bigl(\textstyle\sum_{d,t=1}^{D,T}\displaystyle\frac{\chi_{dt}}{\widetilde{\sigma}^2_d}(y_{dt}-\widetilde{b}_{dk})^2 \nonumber \\
&\hspace{1.95cm}+\widetilde{\cvect}_k\tp\widetilde{\Gammamat}_k^{-1}\widetilde{\cvect}_k-\mvect_k\tp\Vmat_k^{-1}\mvect_k\bigr)\biggr)
\end{align}
where the weights $\{\rho_k\}_{k=1}^K$ are normalized to sum to 1. This formulation is more general than the unique, complete observation case $(T=1,\chivect=\unvect)$ provided by (\ref{eq:JGMM_inverse_map}). The posterior expectation $\mathbb{E}[\xvect|\Yvect_{\chivect}]=\sum_{k=1}^K\rho_k\mvect_k$ can be used to obtain an estimate of the sound source position given an observed spectrogram. Alternatively, one may use the full posterior distribution and, for instance, combine it with other external probabilistic knowledge to increase the localization accuracy or extract higher order information.

\section{Extension to Multiple Sound Sources}
\label{sec:vessl}
We now extend the single sound source localization method described in Section~5 to multiple sound source separation and localization. With our model, this problem can be
formulated as a piecewise affine inversion problem, where observed
signals generated from multiple sources (modeled as latent variables) are both mixed
and corrupted by noise. We extend the PPAM model presented in the previous section to this more general case and propose a variational
expectation-maximization (VEM) approach\cite{Beal03b} to solve for the model inference. The VEM algorithm described
below will be referred to as \textit{variational EM for sound
separation and localization} (VESSL). Typical examples of the algorithm's inputs
and outputs are shown in Fig.~\ref{fig:algo}.

\subsection{The Mixed PPAM Model}
Given a time series of $T$ interaural cues $\Yvect=\{\Yvect_t\}_{t=1}^T\subset\mathbb{R}^D$, we seek the $M$ emitting sound source positions, denoted by
$\Xvect=\{\Xvect_m\}_{m=1}^M\subset\mathbb{R}^L$ ($M$ is assumed to be known). To deal with mixed data, we introduce a \textit{source assignment} variable
$\Wvect=\{W_{dt}\}_{d=1,t=1}^{D,T}$ such that $W_{dt}=m$
when $Y_{dt}$ is generated from source $m$. The only observed data are the interaural cues $\Yvect$ while all the other variables $\Wvect\in \mathcal{W}$, $\Xvect \in \mathcal{X}$ and $\Zvect \in \mathcal{Z}$ are hidden.
To account for $\Wvect$, the observation model rewrites $p(\yvect_t | \wvect_t,\xvect,\zvect) = \textstyle{\prod_{d}} p(y_{dt} | w_{dt}, \xvect_{w_{dt}}, z_{w_{dt}})$
where $p(y_{dt}| W_{dt}=m,\Xvect_m=\xvect_m, Z_m=k) = \mathcal{N}(y_{dt};$ $\avect_{dk}\tp \xvect_m + \bvect_{dk} ,\sigma_d^2)$.
We assume that the different source positions are
independent, yielding $p(\xvect,\zvect)= \textstyle{\prod_{m=1}^M} p(\xvect_{m}, z_{m})$.
Source assignments are also assumed to be independent
over both time and frequency, so that
 $p(\wvect) = \prod_{d,t} p(w_{dt})$. We define the prior on source assignments by $p(W_{dt}=m)=
 \lambda_{dm}$, where $\lambda_{dm}$ are positive numbers representing the relative presence of each source in each frequency channel (source weights), so that
 $\sum_{m=1}^M \lambda_{dm} = 1$ for all $d$. We will write
 $\textstyle{\Lambdamat=\{\lambda_{dm}\}_{d=1,m=1}^{D,M}}$.
Finally, source assignments and positions are assumed independent, so that we get the following hierarchical decomposition of the full model: $p(\Yvect,\Wvect,\Xvect,\Zvect;\psivect)=$
\begin{equation}
p(\Yvect|\Wvect,\Xvect,\Zvect;\psivect)p(\Xvect,\Zvect;\psivect)p(\Wvect;\psivect)
\end{equation}
where
\begin{equation}\psivect = \{\{\Gammamat_k, \cvect_k, \Amat_k,
\bvect_k\}_{k=1}^K, \Sigmamat, \Lambdamat\}
\end{equation}
denotes the complete set of model parameters. 
This extended PPAM model for multiple sound sources will be referred to as the \textit{mixed PPAM} model and is represented with a graphical model in Fig.~\ref{fig:graphical_model}(b).

Notice that the training stage, where position-to-interaural-cue couples $\{(\xvect_n,\yvect_n)\}_{n=1}^N\subset \mathcal{X}\times \mathcal{Y}$ are given (Section~5), 
may be viewed as a particular instance of mixed PPAM where $T=M=N$ and with $\Xvect$ and $\Wvect$ being completely known. Amongst the parameters $\psivect$ of mixed PPAM, the values of $\{\Gammamat_k, \cvect_k, \Amat_k,
\bvect_k\}_{k=1}^K$ have thus already been estimated during this training stage, and only the parameters $\{\Sigmamat, \Lambdamat\}$ remain to be estimated, while $\Xvect$ and $\Wvect$ are now hidden variables. $\Sigmamat$ is re-estimated to account for possibly higher noise levels in the mixed observed signals compared to training.

\subsection{Inference with Variational EM}
From now on, $\mathbb{E}{q}$ denotes the expectation with respect to a
probability distribution $q$. Denoting current parameter values by
$\psivect^{(i)}$, the proposed VEM algorithm  provides, at each
iteration $(i)$, an approximation $q^{(i)}(\wvect,\xvect,\zvect)$ of the
posterior probability $p(\wvect, \xvect, \zvect | \yvect ; \psivect^{(i)})$ that
factorizes as
\begin{equation}
q^{(i)}(\wvect, \xvect, \zvect) = q^{(i)}_W(\wvect)\; q^{(i)}_{X,Z}(\xvect,\zvect)
\end{equation}
where $q^{(i)}_W$ and $q^{(i)}_{X,Z}$ are
probability distributions on $\mathcal{W}$ and $\mathcal{X} \times \mathcal{Z}$
respectively. Such a factorization may seem drastic but its main
beneficial effect is to replace potentially complex stochastic dependencies between latent
variables with deterministic dependencies between relevant moments
of the two sets of variables.
It follows  that the E-step becomes an approximate E-step that can
be further decomposed into two sub-steps whose goals are to
update  $q_{X,Z}$ and $q_W$  in turn. Closed-form expressions for these sub-steps at iteration $(i)$, initialization strategies, and the algorithm termination are detailed below.
\begin{figure*}[t!]
\includegraphics[width = 1\linewidth,clip=,keepaspectratio]{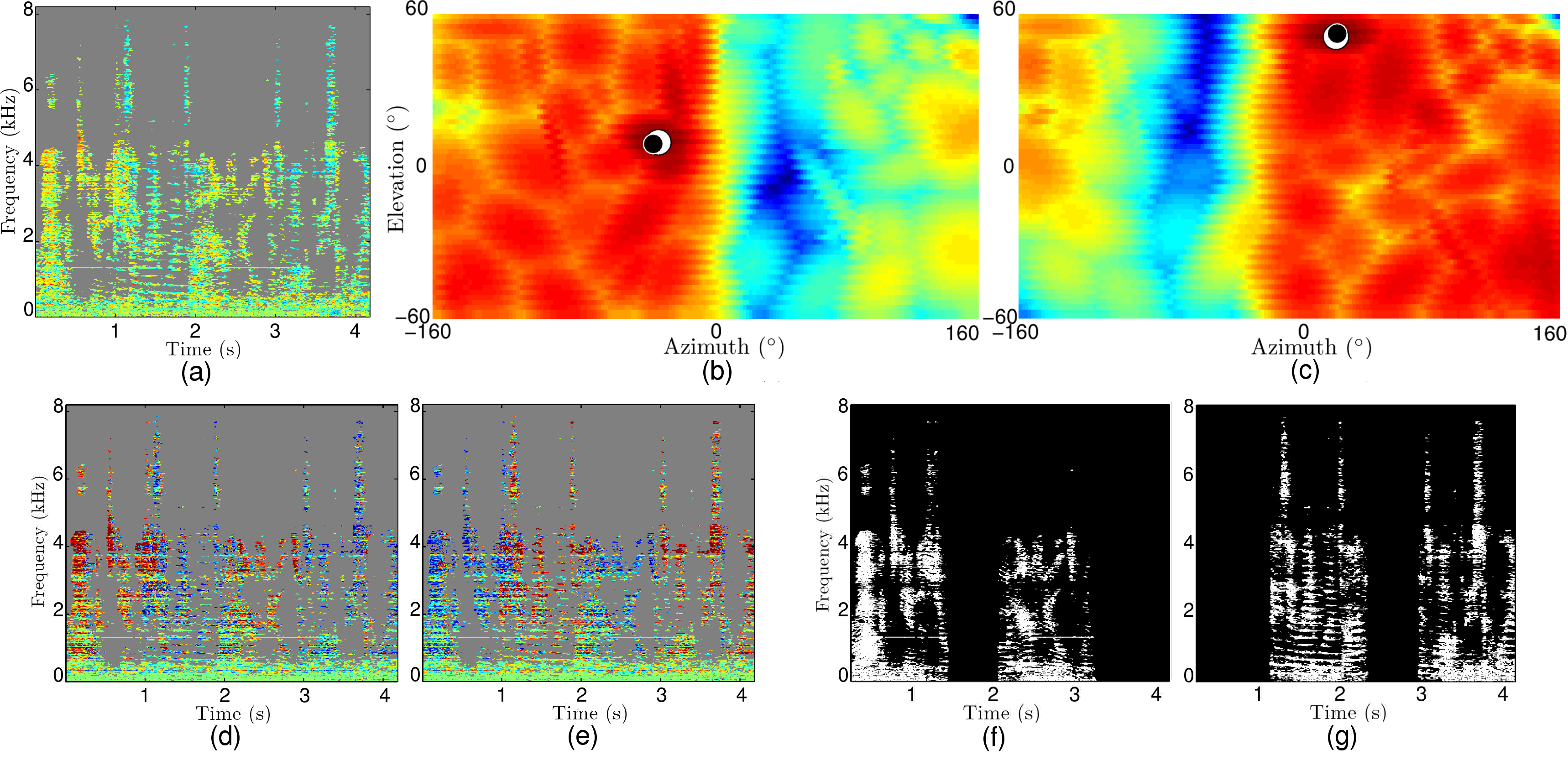}
\vspace{-6mm}
\caption{\small{\label{fig:algo}(a) Input ILD spectrogram. (b,c) Output log-density of each source position as determined by $q^{(\infty)}_{X,Z}$. Ground-truth source positions are noted with a black dot, and the peak of the log-density with a white circle. (d,e) Output source assignment probabilities $q^{(\infty)}_W$. (f,g) Ground truth binary masks. Red color denotes high values, blue color low values, and grey colors missing observations.}}
\vspace{-4mm}
\end{figure*}
\begin{figure*}[!t]
\begin{align}
\label{eq:muS}   
&\muvect_{km}^{(i)} =  \Smat_{km}^{(i)} \bigg(\Gammamat_k^{-1} \cvect_k+ \textstyle\sum_{d,t} \sigma_d^{-2}  q_{W_{dt}}^{(i-1)}(m) (y_{dt}-b_{dk})  \avect_{dk}\bigg), \hspace{2mm}
\Smat_{km}^{(i)} = \bigg(\Gammamat_k^{-1} + \textstyle\sum_{d,t} \sigma_d^{-2}  q_{W_{dt}}^{(i-1)}(m)\avect_{dk} \avect_{dk}\tp\bigg)^{-1},\\
\label{eq:alpha} 
&\alpha_{km}^{(i)} \propto \frac{|\Smat^{(i)}_{km}|^{\frac{1}{2}}}{|\Gammamat_k|^{\frac{1}{2}}}\exp\left\{-\frac{1}{2}\left(\textstyle\sum_{d,t}\displaystyle\frac{\chi_{dt}}{\sigma^2_d}(y_{dt}-\widetilde{b}_{dk})^2
+\cvect_k\tp\Gammamat_k^{-1}\cvect_k-\muvect_{km}^{(i)\top}\Smat^{(i)-1}_{km}\muvect_{km}^{(i)}\right)\right\}, \\
\label{eq:EW-step} 
&q_{W_{dt}}^{(i)}(m) \propto \chi_{dt}\lambda^{(i)}_{dm}\prod_{k=1}^K
\exp\left\{-\frac{\alpha_{km}^{(i)}}{2\sigma_d^2}\left(\operatorname{tr}(\Smat_{km}^{(i)}
\avect_{dk} \avect_{dk}\tp)+ (y_{dt}-\avect_{dk}\tp
\muvect_{km}^{(i)}-b_{dk})^2\right)\right\},\\
\label{eq:Msigma-step} 
&\lambda^{(i)}_{dm} = \frac{1}{T}\textstyle\sum_{t} q^{(i)}_{W_{dt}}(m),\;
\sigma^{2(i)}_{d} =
\displaystyle\frac{\sum_{t,m,k}
q_{W_{dt}}^{(i)}(m)\; \alpha_{km}^{(i)}\;
\bigl(\operatorname{tr}(\Smat_{km}^{(i)} \avect_{dk} \avect_{dk}\tp) + (y_{dt}-
\avect_{dk}\tp\muvect_{km}^{(i)} -b_{dk})^2\bigr)}{
\sum_{t,m,k}
q_{W_{dt}}^{(i)}(m)\; \alpha_{km}^{(i)}}.
\end{align}
\vspace*{-8mm}
\end{figure*}

\noindent
\textbf{E-XZ step:} The update of $q_{X,Z}$ is given by:
$$
q_{X,Z}^{(i)}(\xvect,\zvect) \propto
\exp
\mathbb{E}{q_W^{(i-1)}}[\log p(\xvect,\zvect | \yvect, \Wvect ; \psivect^{(i)})].
$$
It follows from standard algebra that
$$q_{X,Z}^{(i)}(\xvect,\zvect) =\textstyle\prod_{m=1}^M \alpha_{km}^{(i)} \mathcal{N}(\xvect; \muvect_{km}^{(i)}, \Smat_{km}^{(i)})$$ where
$\muvect_{km}^{(i)},\Smat_{km}^{(i)},\alpha_{km}^{(i)}$ are given in (\ref{eq:muS}),(\ref{eq:alpha}) and the weights $\{\alpha_{km}^{(i)}\}_{k=1}^{K}$ are normalized to sum to 1 over $k$ for all $m$.
One can see this step as the \textit{localization step}, since it corresponds to estimating a mixture of Gaussians over the latent space $\mathcal{X}$ of positions for each source. When $M=1$, $\Wvect$ is entirely determined and $q_{W_{dt}}=\chi_{dt}$. Thus, we can directly obtain the probability density $q_{X,Z}$ of the sound source position using (\ref{eq:muS}),(\ref{eq:alpha}), and we recover exactly the single-source formula (\ref{eq:spectrogram_gmm_m}), (\ref{eq:spectrogram_gmm_V}), (\ref{eq:spectrogram_gmm_rho}).

\noindent
\textbf{E-W step:} The update of $q_{W}$ is given by:
$$
q_W^{(i)}(\wvect) \propto
\exp\mathbb{E}{q_{X,Z}^{(i)}}[\log p(\wvect | \yvect, \Xvect, \Zvect;
\psivect^{(i)})].
$$
\black{It follows} that $q_W^{(i)}(\wvect) = \prod_{d,t}
q_{W_{dt}}^{(i)}(w_{dt})$ where $q_{W_{dt}}^{(i)}$ is given in (\ref{eq:EW-step}) and \black{is} normalized to sum to 1 over $m$. This can be seen as the \textit{sound source separation step}, as it provides the probability of assigning each observation to each source. 

\noindent
\textbf{M step:} We need to solve for:
$$\psivect^{(i+1)}  =
\operatorname{argmax}_{\psivect} \mathbb{E}q_W^{(i)} q_{X,Z}^{(i)}[\log
p(\yvect, \Wvect, \Xvect, \Zvect ; \psivect)].
$$
This reduces to the update of the source weights $\Lambdamat^{(i)}$ and noise variances $\Sigmamat^{(i)}=\operatorname{diag}(\sigma_1^{2(i)}...\sigma_D^{2(i)})$ as given by (\ref{eq:Msigma-step}).

\begin{table*}
\centering
   \begin{tabular}{|c|c|c|c|c|c|c|}
      \hline
          Method   & \multicolumn{2}{|c|}{ILD only} &  \multicolumn{2}{|c|}{IPD only} &  \multicolumn{2}{|c|}{ILPD}\\
          \cline{2-7}
          used     & Azimuth & Elevation & Azimuth & Elevation & Azimuth & Elevation \\ \hline
          \cline{2-7}
      PPAM         & {\bf2.2$\pm$1.9}  & {\bf1.6$\pm$1.6} & {\bf1.5$\pm$1.3} & {\bf1.5$\pm$1.4} & {\bf1.8 $\pm$1.6}  & {\bf1.6$\pm$1.5}  \\
      MPLR         & 2.4$\pm$2.2  & 2.2$\pm$2.1  & 1.8$\pm$1.7 & 1.9$\pm$1.7 & 2.2 $\pm$2.0  & 2.1$\pm$2.0   \\
      SIR-1        & 41$\pm$34   & 17$\pm$13 & 36$\pm$25 & 24$\pm$17 & 41$\pm$34   & 16$\pm$13   \\
      SIR-2        & 27$\pm$28 & 14$\pm$13 & 32$\pm$23 & 11$\pm$11 & 26$\pm$28  & 14$\pm$13   \\
      \hline
   \end{tabular}
   \caption {\small{\label{tab:mapping} Comparing the average and standard deviation (Avg$\pm$Std) of azimuth and elevation angular errors in degrees using different types of interaural vectors obtained from white noise recordings and different mapping techniques.}}
\end{table*}

\noindent
\textbf{Initialization strategies:}
Extensive real world experiments show that the VEM objective function,
called the variational free energy, have a large number of local
maxima. This may be due to the
combinatorial size of the set of all possible binary masks
$\mathcal{W}$ and the set of all possible affine transformation
assignments $\mathcal{Z}$. Indeed, the procedure turns out to be more
sensitive to initialization and to get trapped in
suboptimal solutions more often as the size of the spectrogram and
the number of \black{transformations} $K$ are increased. On the other hand, too
few local affine transformations $K$ make the mapping very
imprecise. We thus developed a new efficient way to deal with
the well established local maxima problem, referred to as
\textit{multi-scale initialization}. The idea is to train PPAM at different \textit{scales},
\textit{i.e.}, with a different number of \black{transformations} $K$ at each scale, yielding different sets of trained parameters
$\widetilde{\theta}_K$ where, \textit{e.g.}, $K=1,2,4,8\dots,64$.
When proceeding with the inverse mapping, we first run the VEM
algorithm from a random initialization using
$\widetilde{\theta}_1$. We then use the obtained masks and
positions to initialize a new VEM algorithm using
$\widetilde{\theta}_2$, then $\widetilde{\theta}_4$, and so
forth until the desired value for $K$. To further improve the
convergence of the algorithm at each scale, an additional
constraint is added, namely \textit{progressive masking}.
During the first iteration, the mask of each source is constrained
such that all the frequency bins of each time frame are assigned
to the same source. This is done by adding a product over $t$ in the expression of
$q^{(1)}_{W_{dt}}(m)$ (\ref{eq:EW-step}). Similarly to what is done in\cite{MandelWeissEllis10}, this constraint is then
progressively released at each iteration by dividing time frames
in $2,4,8\dots,F$ frequency blocks until the total number of frequency bins $F$ is reached.
Combining these two strategies dramatically increases both the performance and
speed\fnm{d}\fnt{d}{About 15 times real time speed for a mixture of 2 sources and $K=64$ using a MATLAB implementation on a standard PC.} of the proposed VESSL algorithm.

\noindent
\textbf{Algorithm termination:}
Maximum a posteriori (MAP) estimates for missing data can be easily obtained at convergence of the algorithm by
maximizing respectively the final probability distributions $q^{(\infty)}_{X,Z}(\xvect,
\zvect)$ and $q^{(\infty)}_W(\wvect)$.
We have $(\Xvect_m^{MAP}, Z_m^{MAP})= (\muvect^{(\infty)}_{\hat{k}m}, \hat{k})$
where $\hat{k} = \operatorname*{argmax}_k \alpha^{(\infty)}_{km}\;
|\Sigmamat^{(\infty)}_{km}|^{-1/2}$ and $W_{dt}^{MAP}=\operatorname*{argmax}_m q^{(\infty)}_{W_{dt}}(m)$.
Note that as shown in Fig.~\ref{fig:algo}, the algorithm not only provides MAP estimates, but also complete posterior distributions over both the
2D space of sound source positions $\mathcal{X}$ and the space of source assignments $\mathcal{W}$. Using
the final assignment probabilities $q^{(\infty)}_{W_{dt}}(m)$ of source $m$, one can
multiply recorded spectrogram values $s_{ft}^{(\textrm{L})}$ and $s_{ft}^{(\textrm{R})}$ by $1$ if $m$ has the highest emission probability at $(f,t)$, and $0$ otherwise (binary masking).
\black{This approximates} the spectrogram $\{s^{(\textrm{m})}_{ft}\}_{f,t=1}^{F,T}$ emitted by source $m$, \black{from which the original temporal signal can be recovered} using inverse Fourier transform, hence achieving sound source separation.

\begin{figure*}[t!]
   \begin{minipage}[c]{.48\linewidth}
      \includegraphics[width = \linewidth,clip=,keepaspectratio]{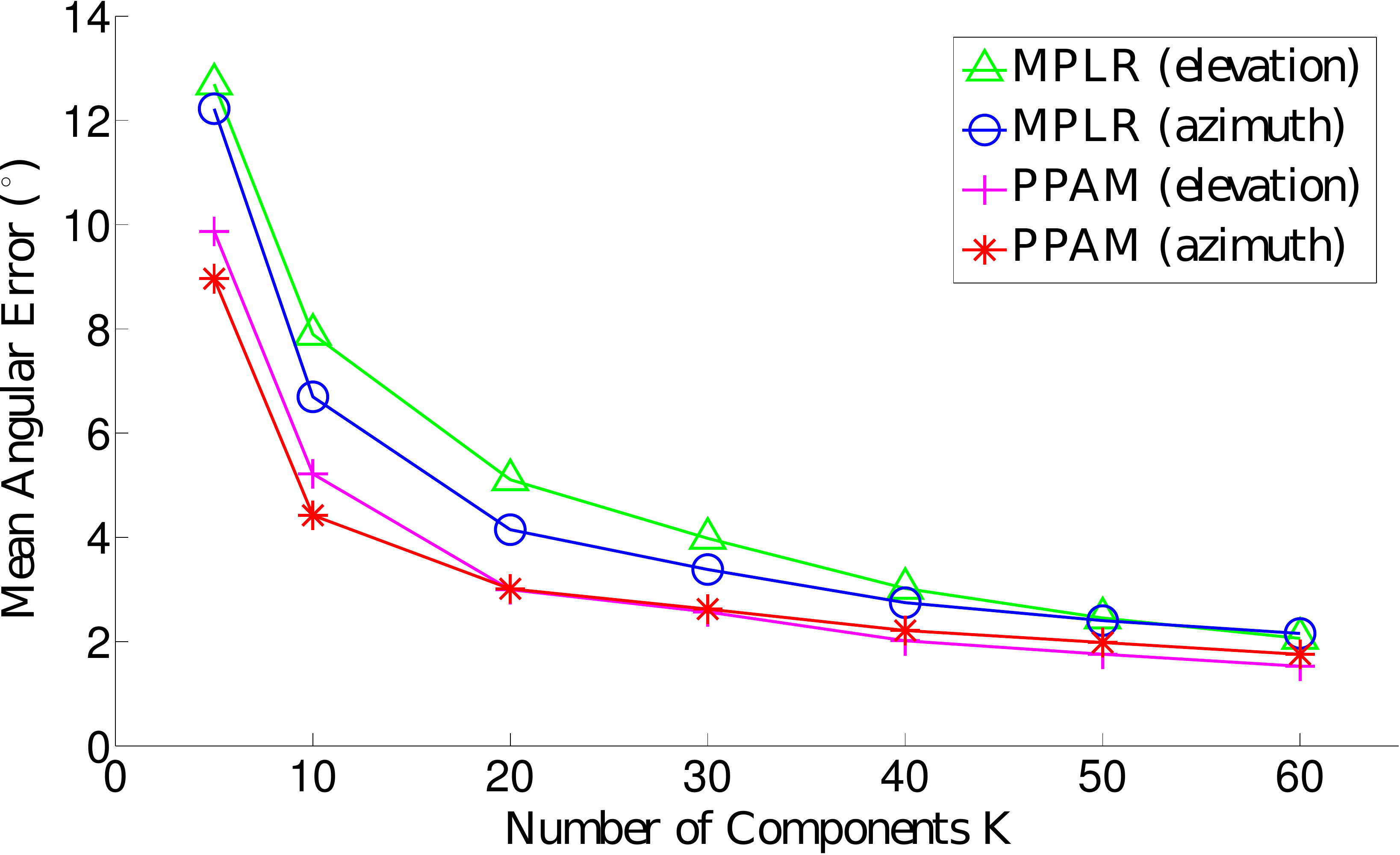}
      \caption{\small{\label{fig:plot_mapping_err} Comparison of the proposed method (PPAM) with MPLR (Qiao and Minematsu 2009) as a function of the number of components.}}
   \end{minipage} \hfill
   \begin{minipage}[c]{.48\linewidth}
      \includegraphics[width = \linewidth,clip=,keepaspectratio]{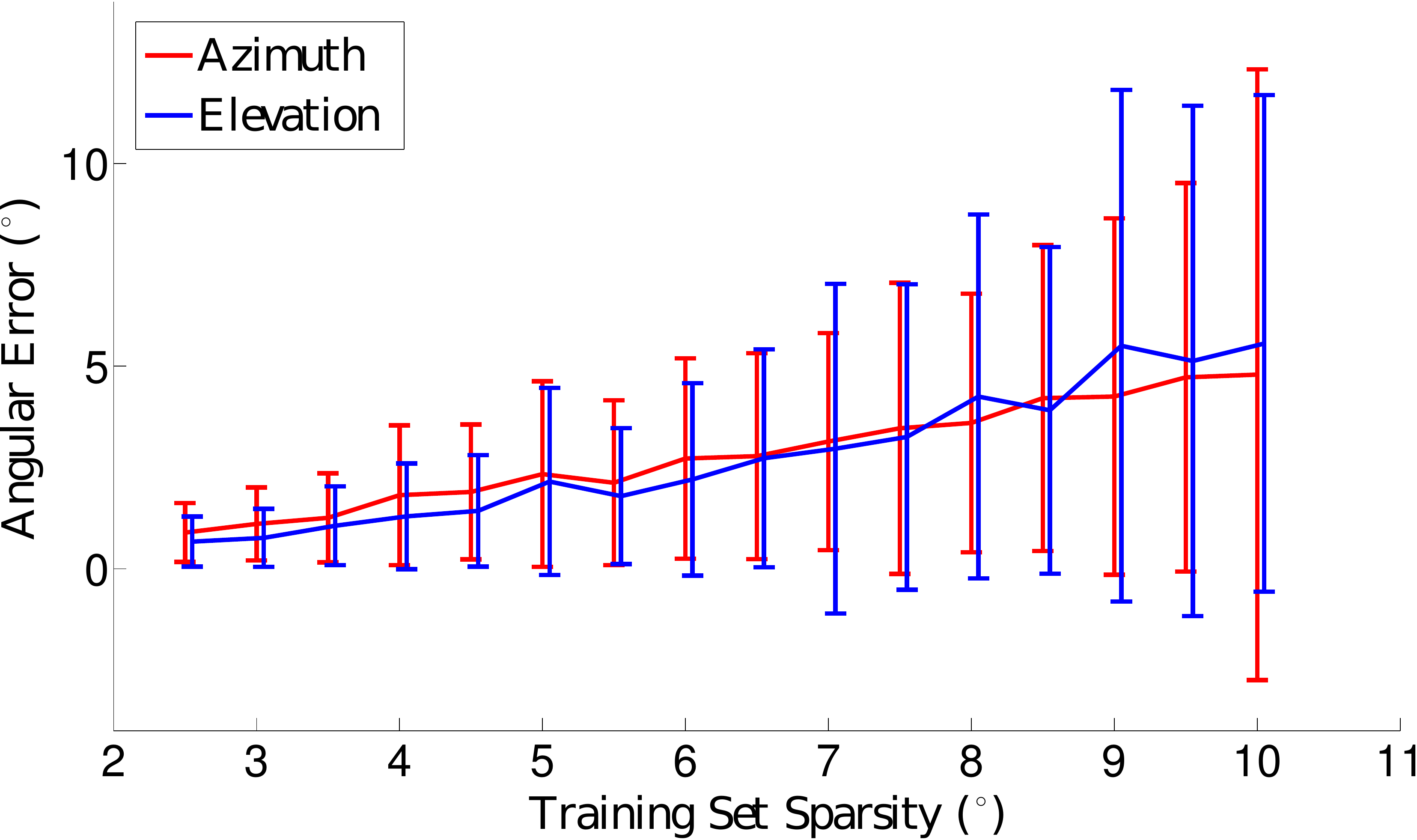}
      \caption{\small{\label{fig:plot_sparsity} Average and standard deviation of angular errors in white noise localization as a function of the training set's sparsity.}}
   \end{minipage}
\end{figure*}
\section{Results on Acoustic Space Mapping, Source Localization and Sound Separation}
\label{sec:results}
We first compare PPAM to two other existing mapping methods, namely mixture of probabilistic linear regression (MPLR\cite{QiaoMinematsu09}), and sliced inverse regression (SIR\cite{Li91}). 
MPLR \black{may be viewed} as a unifying view of joint GMM mapping techniques, which consists of estimating a standard Gaussian mixture model on joint variables $[\Xvect;\Yvect]$ and uses conditional expectations to infer the mapping. Joint GMM has been widely used in audio applications\cite{KainMacon98,StylianouMoulines98,TodaBlackTokuda08,QiaoMinematsu09}.
SIR  quantizes the  low-dimensional data $\Xvect$ into {\it
slices} or clusters which in turn induces a quantization of the
$\Yvect$-space. Each $\Yvect$-slice (all points $\yvect_n$ that
map to the same $\Xvect$-slice) is then replaced with its mean and
PCA is carried out on these means. \black{The resulting} low-dimensional representation is then informed by $\Xvect$ values through the
preliminary slicing. We selected one (SIR-1) or two (SIR-2) principal axes for
dimensionality reduction, 20 slices (the number of slices is known
to have little influence on the results), and polynomial
regression of order three (higher orders did not show significant improvements).  We use $K=64$ for both PPAM and MPLR. All techniques are trained on three types of cues:
full spectrum ILD only, low-IPD only and ILPD cues (see Section~2~
and~4).~
For each type of cue, the training is done on $10$ random subsets of $4,800$ mean interaural vectors obtained from white noise recordings emitted by sources spanning a $[-160^\circ,160^\circ]$ azimuth range and $[-60^\circ,60^\circ]$ elevation range (section~3),~
and tested on the mean interaural vectors of the remaining $4,800$ source positions ($96,000$ tests in total). The resulting angular errors are \black{shown} in Table \ref{tab:mapping}. As it may be seen, PPAM always performs best, validating the choice of piecewise-affine probabilistic technique adapted to the manifold structure of interaural data. The poor performance of SIR can be explained by the important non-linear nature of the data. Although pure IPD cues show slightly better localization results than pure ILD cues, it is better to use both to maximize the available information in the case of noisy and/or missing data. Other experiments in this section are therefore carried out using ILPD cues.

Fig.~\ref{fig:plot_mapping_err} shows the influence of the number of affine components $K$ on PPAM's and MPLR's localization results using similar test and training sets. Better performance of PPAM can be explained by the fact that the number of parameters to estimate in joint GMM methods increases quadratically with the dimensionality, and thus becomes prohibitively high using ILPD data ($D=730$). Unsurprisingly, the localization error decreases when $K$ increases. In practice, too high values of $K$ (less than 20 samples per affine component) may lead to some degenerate covariance matrices in components where there are too few samples. Such components are simply removed along the execution of the algorithm, thus reducing $K$. In other words, the only free parameter $K$ of PPAM can be chosen based on a compromise between computational cost and precision.
\begin{Table}
   \center
   \begin{tabular}{|c|c|c|}
      \hline
       Method   & \multicolumn{2}{|c|}{1 source} \\
       \cline{2-3}
       used           & Az $({}^\circ)$ & El $({}^\circ)$ \\
       \hline
       PPAM(\textit{T1})& {\boldmath $2.1\pm 2.1$} & {\boldmath $1.1\pm 1.2$}\\
       \hline
       PPAM(\textit{T2})& $3.5\pm 3.3$    & $2.4\pm 2.6$     \\
       \hline
       PHAT             & $7.6\pm 9.4$    & \\
       \hline
   \end{tabular}
   \captionof{table}{\small{\label{tab:locresults} \black{Average single sound source localization error in degrees using PHAT and PPAM with two different training sets.}}}
\end{Table}

Fig.~\ref{fig:plot_sparsity} shows the influence of the training set's \textit{sparsity} on PPAM's localization results.  We represent the sparsity by an angle $\delta$, corresponding to the average pan and tilt absolute difference between a point and its nearest neighbor in the training set. The complete training set of sparsity $2^{\circ}$ was uniformly decimated at random to obtain irregularly spaced and smaller training sets, while the test positions were randomly chosen from \black{the} remaining ones. For a given sparsity $\delta$, 10 different decimated sets were used for training, and 100 source positions were estimated for each one, \textit{i.e.}, 1000 localization tasks. $K$ was chosen to have approximately 30 training samples per affine component. The mean and standard deviation of the errors in azimuth and in elevation are shown in Fig.~\ref{fig:plot_sparsity}. The mean localization errors are always smaller than half the training set sparsity, which illustrates the interpolation ability of our method. In addition, the error's standard deviation remains reasonable even for heavily decimated training sets ($384$ points only for $\delta=10^\circ$), thus showing that the overall performance is not much affected by the distribution and size of the training set being used. With the automatic audio-motor technique described in Section~3,~
recording a training set of sparsity $2^{\circ}$ ($9,600$ points) requires 2 hours and 30 minutes while it takes 25 minutes with a sparsity of $5^{\circ}$ ($1,530$ points).

\begin{table*}
\centering
   \begin{tabular}{|c|c|c|c|c|c|c|c|c|}
      \hline
          Method   & \multicolumn{4}{|c|}{2 sources} &  \multicolumn{4}{|c|}{3 sources}\\
          \cline{2-9}
          used     & Az & El & SDR & STIR & Az & El & SDR & STIR  \\
          \hline
      VESSL(\textit{T1})     & {\bf4.7$\pm$11}  & {\bf2.9$\pm$9.9} & {\bf3.8$\pm$1.7} & {\bf6.1$\pm$1.7}
                             & {\bf12 $\pm$21}  & {\bf8.7$\pm$19}  & {\bf1.7$\pm$1.5} & {\bf2.2$\pm$1.5} \\
      VESSL(\textit{T2})     & 8.2$\pm$16  & 4.7$\pm$11  & 3.3$\pm$1.6 & 5.2$\pm$1.6
                             & 16 $\pm$24  & 9.1$\pm$18  & 1.5$\pm$1.5 & 1.8$\pm$1.5 \\
      MESSL-G      & 14$\pm$21   &             & 2.3$\pm$1.6 & 6.0$\pm$4.3
                   & 18$\pm$28   &             & 1.3$\pm$1.2 & 2.1$\pm$4.4 \\
      mixture      &             &             & 0.0$\pm$2.5 & 0.2$\pm$2.5
                   &             &             & -3.2$\pm$2.3& -3.0$\pm$2.3 \\
      oracle       &             &             & 12$\pm$ 1.6 & 21 $\pm$2.0
                   &             &             & 11$\pm$ 1.7 & 20 $\pm$2.1 \\
      \hline
   \end{tabular}
   \caption {\small{\label{tab:results} Comparing the average and standard deviation (Avg$\pm$Std) of azimuth (Az) and elevation (El) angular errors in degrees, as well as Signal to Distortion Ratio (SDR) and Signal to Inteferer Ratio (STIR) for 600 test mixtures of 2 to 3 sources using different methods.}}
\end{table*}

We also tested the ability of PPAM to localize a single sound source using real world recordings of randomly located sources emitting random utterances (see Section~3),~
using the technique presented in Section~5.2.~
We use both the complete set of ILPD cues \textit{T1} (sparsity $\delta=2^{\circ}$) and a partial set \textit{T2} of sparsity $\delta=5^{\circ}$ for training. We set $K=128$ for \textit{T1} and $K=64$ for \textit{T2}. \black{Test sounds are chosen to be outside of the training set \textit{T2}.} The spectral power threshold is manually set quite high, so that test spectrograms has $89.6\%$ of missing data, \black{on average}. Results are compared to a baseline sound localization method, namely PHAT-histogram\cite{Aarabi02}. As the vast majority of existing source localization methods\cite{YilmazRickard04,MoubaMarchand06,MandelEllisJebara07,VisteEvangelista03,LiuWang10,WoodruffWang12}, PHAT-histogram provides a time difference of arrival (TDOA) which can only be used to estimate frontal azimuth. The link between TDOAs and azimuths is estimated using linear regression on our dataset. For the comparison to be fair, PHAT is only tested on frontal sources. Comparison with the few existing binaural 2D sound source localization methods\cite{KullaibAlmuallaVernon09,KeyrouzMaierDiepold07} is not possible because they rely on additional knowledge of the system being used that are not available. Means and standard deviations of azimuth (Az) and elevation (El) errors (Avg$\pm$Std) are shown in Table \ref{tab:locresults}.

\noindent
The proposed acoustic space learning approach dramatically outperforms the baseline.
No front-back azimuth or elevation confusions is observed using our method, thanks to the asymmetry of the dummy head and to the spatial richness of interaural spectral cues.

We finally tested the ability of VESSL (section~6)~
to localize and separate multiple sound sources emitting at the same time. We use both training sets \textit{T1} and \textit{T2} for training and
mixtures of 2 to 3 sources for testing\fnm{e}\fnt{e}{\black{The binaural algorithms considered performed equally poorly on mixtures of 4 sources or more.}}. Mixtures are obtained
by summing binaural recordings of sources emitting random utterances from random positions (see Section~3)~
so that at least two sources were emitting at the same time in
$60\%$ of the test sounds. 
Localization and separation results are compared to state-of-the-art EM-based sound source
separation and localization algorithm MESSL\cite{MandelWeissEllis10}. MESSL \black{does} not rely on acoustic space learning, and
estimates a time difference of arrival for each source. As for PHAT-histogram, results given for MESSL correspond to tests with frontal
sources only. The version MESSL-G that is used includes a garbage component and ILD priors
to better account for reverberations and is reported to outperform
four methods in reverberant conditions in terms of separation\cite{YilmazRickard04,BuchnerAichnerKellermann05,MoubaMarchand06,SawadaArakiMakino07}. We evaluate
separation performance using the standard metrics signal to
distortion ratio (SDR) and signal to interferer ratio (STIR)
introduced in\cite{VincentGribonval06-short}. SDR and STIR scores
of both methods are also compared to those obtained with the
ground truth binary masks or \textit{oracle masks}\cite{YilmazRickard04} and to those of the original mixture.
Oracle masks provide an upper bound for binary masking methods that cannot be reached in
practice because it requires knowledge of the original signals. Conversely,
the mixture scores provide a lower bound, as no mask is applied. Table \ref{tab:results} shows that VESSL significantly outperforms MESSL-G both in terms of separation and localization, putting forward
acoustic space learning as a promising tool for accurate sound source localization and separation.

\section{Conclusion and Future Work}
\label{sec:conclusion}
We showed the existence of binaural manifolds, \textit{i.e.} an intrinsic locally-linear, low-dimensional structure hidden behind the complexity of interaural data obtained from real world recordings. Based on this key property, we \black{developed} a probabilistic framework to learn a mapping from sound source positions to interaural cues. We showed how this framework could be used to accurately localize in both azimuth and elevation and separate \black{mixtures} of natural sound sources. Results show the superiority of acoustic space learning compared to other sound source localization techniques relying on simplified mapping models.

In this work, auditory data are mapped to the motor-states of a robotic system, but the same framework could be exploited with mappings of different nature. \black{Typically, for \textit{unsupervised} sound source localization where one only has access to non-annotated auditory data from different locations, one may map auditory cues to their intrinsic coordinates obtained using manifold learning, as explained in Section~4.}~
Such coordinates are spatially consistent, \textit{i.e.}, two sources that are near will yield near coordinates, but are not linked to a physical quantity and may therefore be hard to interpret or evaluate. Alternatively, auditory data could be annotated with the pixel position of the sound source in an image, yielding audio-visual mapping instead of audio-motor mapping.

A direction for future work is to study the influence of changes in the reverberating properties of the room where the
training is performed. The PPAM model could be extended by adding latent components modeling such changes\cite{DeleforgeHoraud13b}, thus becoming more robust to the recording environment. Another 
open problem is determining the number of sound sources $M$ in the mixture, which is generally challenging in source separation. In our framework, it corresponds to a model selection problem, which could be addressed using, \textit{e.g.}, a Bayesian information theoretic analysis.

\nonumsection{References}

\bibliographystyle{my_ijns}

\end{multicols}
\end{document}